\documentclass[twocolumn]{aastex701}
\usepackage{natbib}
\usepackage{xspace}

\newcommand{\squiggle}{SQuIGG$\vec{L}$E \,}
\newcommand{\squigglecomma}{SQuIGG$\vec{L}$E}

\newcommand{\qsorate}{$4.6^{+1.6}_{-1.2}\%$\xspace}

\newcommand{\oiii}{[O{\scshape iii}]\xspace}

\newcommand{\nii}{[N{\scshape ii}]\xspace}

\newcommand{\oii}{[O{\scshape ii}]\xspace}

\newcommand{\nev}{[Ne{\scshape v}]\xspace}

\newcommand{\halpha}{H$\alpha$\xspace}
\newcommand{\hbeta}{H$\beta$\xspace}

\defcitealias{soto2025}{S25}

\begin{document}

\title{Extreme MgII Emission in Recently-Quenched Galaxies: A New Sample of Intermediate-Redshift Post-Starburst Quasars from the DESI Early Data Release}

\author[0000-0003-1535-4277]{Margaret E. Verrico}\thanks{Email: verrico2@illinois.edu}
\affiliation{University of Illinois Urbana-Champaign Department of Astronomy, University of Illinois, 1002 W. Green St., Urbana, IL 61801, USA}
\email{verrico2@illinois.edu}

\author[0009-0006-5774-8636]{Aayan Verma}
\affiliation{University of Illinois Urbana-Champaign Department of Astronomy, University of Illinois, 1002 W. Green St., Urbana, IL 61801, USA}
\email{aayanverma2005@gmail.com}

\author[0000-0002-4235-7337]{K. Decker French}
\affiliation{University of Illinois Urbana-Champaign Department of Astronomy, University of Illinois, 1002 W. Green St., Urbana, IL 61801, USA}
\email{deckerkf@illinois.edu}

\author[0000-0003-4075-7393]{David J. Setton}\thanks{NHFP Hubble Fellow}
\affiliation{The William H. Miller III Department of Physics and Astronomy, The Johns Hopkins University, Baltimore, MD 21218}
\affiliation{Department of Astrophysical Sciences, Princeton University, 4 Ivy Lane, Princeton, NJ 08544, USA}
\email{davidsetton@jhu.edu}

\author[0000-0001-9328-3991]{Omar Almaini}
\affiliation{School of Physics and Astronomy, University of Nottingham, University Park, Nottingham NG7 2RD, UK}
\email{omar.almaini@nottingham.ac.uk}

\author[0000-0002-5612-3427]{Jenny E. Greene}
\affiliation{Department of Astrophysical Sciences, Princeton University, Princeton, NJ 08544, USA}
\email{jennyg@princeton.edu}

\author[0000-0002-5907-3330]{Stephanie LaMassa}
\affiliation{Space Telescope Science Institute, 3700 San Martin Drive, Baltimore, MD 21218, USA}
\email{slamassa@stsci.edu}

\author[0000-0002-0696-6952]{Yuanze Luo}
\affiliation{Department of Physics and Astronomy and George P. and Cynthia Woods Mitchell Institute for Fundamental Physics and Astronomy, Texas A\&M University, 4242 TAMU, College Station, TX 77843-4242, US}
\email{yluo37@tamu.edu}

\author[0000-0002-9471-8499]{Pallavi Patil}
\affiliation{National Radio Astronomy Observatory, 520 Edgemont Road, Charlottesville, VA 22903-2475, USA}
\email{ppatil@nrao.edu}

\author[0000-0001-7883-8434]{Kate Rowlands}
\affiliation{AURA for ESA, Space Telescope Science Institute, 3700 San Martin Drive, Baltimore, MD 21218, USA}
\affiliation{William H. Miller III Department of Physics and Astronomy, Johns Hopkins University, Baltimore, MD 21218, USA}
\email{krowlands@stsci.edu}

\author[0009-0006-5194-1777]{Dharlenny Soto}
\affiliation{Department of Astrophysics, CUNY Graduate Center, 365 5th Ave, New York, NY 10016}
\email{lsoto2440@gmail.com}

\begin{abstract}

Post-starburst galaxies are galaxies which have recently and rapidly shut off star formation. AGN activity is often invoked to explain this rapid quenching, but most post-starburst selection methods select against optical AGN, motivating searches for ``post-starburst quasars." We present DESI observations of ten post-starburst quasars with significant MgII emission ($L_{\text{MgII}} \geq 3 \times 10^{41}$ erg/s). This is the first sample of post-starburst quasars to be selected from a post-starburst galaxy parent sample rather than a quasar catalog. Quasars make up \qsorate of the post-starburst parent sample at $0.49 \leq z \leq 1.39$. Using multiwavelength AGN diagnostics, we find that $26^{+3.0}_{-2.8}\%$ of post-starbursts in the parent sample host AGN. We find that post-starburst selection methods are biased against AGN with $\text{L}_{\text{Bol}}\gtrsim10^{45} \textrm{erg s}^{-1}$, potentially removing the quasars responsible for driving outflows that aid in quenching. Finally, we find that 9/10 post-starburst quasars in our sample have no detected broad \hbeta, similar to so-called ``MgII emitters" that may be the remnants of fading quasars; if confirmed, these objects suggest a link between post-starburst galaxies and variable AGN. Our new sample of post-starburst quasars provides a link between previously reported local and high redshift post-starburst quasars during the epoch when star formation throughout the universe was declining.

\end{abstract}

\keywords{\uat{Post-starburst galaxies}{2176}, \uat{Active galactic nuclei}{16}, \uat{Galaxy quenching}{2040}}

\section{Introduction}

Massive galaxies in the local Universe can be broadly divided into two categories: star-forming spiral galaxies like the Milky Way and non-star-forming quiescent ellipticals. The existence of these two types of galaxy imply that some process occurs to end, or ``quench," star formation in galaxies \citep[see][for a recent review]{whitaker2026}. There is evidence that quenching can happen slowly \citep[e.g.][]{Martig2009,fabian2012,peng2015} or quickly \citep[e.g.][]{hopkins2008}, with slow quenching mechanisms dominating in the local Universe \citep[][]{schawinski2014,peng2015,park2022} and rapid quenching being dominant at high redshift \citep[][]{setton2023,park2024}. It is the rapid quenching mode that formed as much as half of the massive quiescent galaxy population at $z=0$ \citep{wild2009,snyder2011,wild2016}; to study this mode, we need to select galaxies that have recently shut down star formation.

Recently and rapidly quenched galaxies are often called ``post-starburst" and are characterized by a dominant A-star population formed in the past $\lesssim1$ Gyr with little or no ongoing star formation (\citealt{dressler1983,couch1987,Zabludoff1996,Goto2005}, and see review by \citealt{French2021}). Post-starbursts were traditionally selected based on their large Balmer breaks and lack of \halpha and/or OII emission \citep[][]{dressler1983,Zabludoff1996}, though techniques that allow for more emission from shocks and AGN \citep[][]{Alatalo2016a}, dimension-reduction techniques like Principal Component Analysis \citep[][]{Wild2007,wild2009}, and color selections to be used with photometric surveys have been developed \citep[e.g.][]{kriek2010,whitaker2012a,wild2014}. 

To shut down star formation, cold molecular gas---the fuel for star formation---must be removed or rendered less efficient at forming stars. Several rapid mechanisms for star formation suppression during the post-starburst phase have been proposed. One potential mechanism is supermassive black hole accretion in the form of an ``active galactic nucleus," or AGN.  Energy from AGN accretion is thought to couple to gas in the galaxy in a process known as``AGN feedback" \citep[][]{silk1998,dimatteo2005,fabian2012}, removing it via winds and outflows \citep[e.g.][]{king2003,king2010,kingandpounds2015} or lowering its star formation efficiency by destroying molecular gas clumps and heating gas \citep[e.g.][]{ellison2021,garciaburillo2024,esposito2026}. On larger scales, AGN can heat or drive cavities in the circumgalactic medium, preventing new gas infall onto galaxies in clusters \citep[e.g.][]{fabian2006,forman2007}. In simulations, AGN feedback is required to fully shut down, or ``quench," star formation in massive galaxies and form the massive quiescent galaxy population observed at low redshift \citep[e.g.][]{EAGLESIM,kaviraj2017,TNGSIM,scharre2024}. Observational results on the correlation between star formation and AGN activity are mixed \citep[e.g.][]{chen2013,aird2019,Cristello2024,mountrichas2024}, complicated by the difficulty in measuring AGN host galaxy properties \citep[e.g.][]{cardoso2017} and rapid AGN variability on timescales much shorter than those relevant for star formation \citep[e.g.][]{novak2011,hickox2014,Schawinski2015}.

AGN feedback in post-starbursts is of particular interest, as the AGN may aid in removing residual gas after a galaxy merger-triggered starburst \citep{hopkins2008}. Quasar-driven wind launching occurs at Eddington or super-Eddington accretion rates \citep{kingandpounds2015} and should clear the environment around the AGN, leading to an unobscured ``post-starburst quasar" phase in young post-starburst galaxies \citep{kocevski2015}. Evidence of past AGN episodes has been found in post-starbursts in the form of extended emission line regions \citep[EELRs; see e.g.][]{lintott2009,greene2011,keel2012,Prieto2016,french2023} and outflows \citep[e.g.][]{alatalo2011,fodor2025,luo2026b}, but post-starbursts hosting even moderately luminous AGN are traditionally removed from post-starburst samples due to their emission and blue continuum which mimic ongoing star formation. Based on narrow-line diagnostics and multiwavelength AGN criteria, estimates range from $\sim$a few percent \citep[e.g.][]{greene2020,luo2026,patil2026} to as many as half \citep[e.g.][]{brown2009,pawlik2018,skarbinski2026} of post-starbursts hosting AGN, with selection method and redshift range changing the measured fraction.

Motivated by the bias against AGN in post-starburst selection methods, adapted selection methods have been employed to search for ``post-starburst quasars," or post-starbursts hosting luminous/broad-line AGN \citep[][]{brotherton1999}. Studies of large spectroscopic samples from ground-based telescopes have yielded samples of post-starbursts with broad-line AGN at $z \leq 0.5$ \citep[][]{cales2011,melnick2015,wei2018}. These objects are some of the best opportunities to identify AGN feedback in action in post-starbursts, but as they were all selected from quasar catalogs, they cannot be used to estimate the overall quasar rate among post-starbursts. As these samples were selected from low-redshift galaxies, they also do not probe the high-redshift processes that formed the massive quiescent galaxies we see today \citep{macleod2016,weaver2023}, motivating a search for post-starburst quasars at Cosmic Noon when the red sequence formed. Small samples of $z>1.5$ post-starburst quasars have now been identified using the James Webb Space Telescope \citep[e.g.][]{onoue2025,valentino2026}. However, this redshift range is difficult to observe from the ground, and space-based instruments like the Hubble and James Webb Space Telescopes have small observing areas, which are not optimal for searching for galaxies in the short transition between star formation and quiescence. Intermediate redshift ($0.5 < z < 1.5$) galaxies are therefore a more promising avenue to study galaxy quenching with ground-based survey instruments while still probing similar processes to those that quenched local massive ellipticals.

In this study, we identify ten post-starburst quasars in the \citet{soto2025} catalog of DESI Early Data Release \citep[][]{desi_edr,fastspecfit} post-starburst galaxies at $0.49 \leq z \leq 1.49$. This is the first sample of post-starburst quasars to be selected from a post-starburst galaxy parent sample rather than a quasar catalog. These objects may be the smoking gun for AGN-driven quenching, just at the epoch when star formation throughout the Universe is falling from its peak. We describe the data used in this paper in Section \ref{sec:data}. In Section \ref{sec:selection}, we describe our quasar selection. In Section \ref{sec:results}, we discuss the rate of quasars/AGN in the \citet[]{soto2025} sample and how this rate varies based on selection method (e.g. broad-line selected versus narrow-line-selected). In Section \ref{discussion:fraction}, we compare the multiwavelength AGN fractions of the \citet[]{soto2025} sample to other post-starburst galaxy samples from the literature. We further describe how differences in post-starburst selection method impact the measured AGN rate in Section \ref{discussion:selectionmethods}. In Section \ref{discussion:fading}, we compare our post-starburst quasars to the unusual ``MgII emitters" first identified in \citet[]{roig2014} and theorized to be fading AGN in \citet[]{guo2020}.

Throughout this work, we assume a flat $\Lambda$CDM cosmology with $H_0 = 70$ km/s/Mpc and $\Omega_m = 0.3$.

\section{Data \& Methods} \label{sec:data}

 All data are taken from the Dark Energy Spectroscopic Instrument (DESI) Survey Early Data Release \citep[][]{desi_edr}. Initial post-starburst quasar candidates were selected from the post-starburst galaxy catalog of \citet{soto2025} \citepalias[hereafter][]{soto2025}. We use the published MgII fluxes and redshifts from the Fuji Value Added Catalog published as part of the DESI Early Data Release for quasar selection \citep[][]{desi_edr}. This catalog was produced using the FastSpecFit\footnote{https://fastspecfit.readthedocs.io/en/latest/index.html} stellar continuum and emission line modeling software \citep[][J. Moustakas et al. in preparation]{fastspecfit}. FastSpecFit simultaneously models DESI photometry and spectroscopy to obtain emission line fluxes, continuum luminosities, and estimates of galaxy properties like stellar mass and star formation rate. FastSpecFit uses the C3K stellar library \citep[][]{conroy2012} with the MIST isochrones \citep[][]{dotter2016, choi2016, paxton2011, paxton2013, paxton2015} and assumes a \citet[]{chabrier2003} IMF when generating SPS model spectra. SPS models are generated using five age bins of constant SFR and varying width from 30 Myr to 13.7 Gyr according to the \texttt{Prospector} \texttt{adjust\_continuity\_agebins} SFH template \citep[][]{leja2019}. Stellar metallicity is fixed to solar, dust emission according to the \citet{draine2007} model is fixed ($Q_{\text{PAH}}=3.5\%, U_{\text{min}}=1.0, \gamma=0.01$), and a \citet[]{nenkova2008} AGN torus model with $\tau=10$ is used in fitting. Hydrogen lines are fit with separate broad and narrow components; other emission lines are fit with a narrow component with tied velocity dispersions and velocity shifts.
 
\citetalias{soto2025} selects post-starburst galaxies from the Fuji catalog at $0.49 \leq z \leq 1.39$ based on the following criteria:

\begin{enumerate}
    \item $\text{[OII]} \lambda3727$ equivalent width (EW) $< 3 \textrm{\AA}$, to select for quiescence \citep[e.g.][]{Zabludoff1996,yan2006};
    \item $4$  \AA\ $<$ H$\delta_A <10$ \AA, to select for recent quenching \citep[e.g.][]{dressler1999,brown2009,french2015}; and
    \item $\sigma_{\text{H}\delta} < 2$ \AA\ and $z-$band signal-to-noise ratio (SNR) $>10$, for data quality,
\end{enumerate}

where $\text{[OII]} \lambda3727$ EW and $z-$band SNR were taken from the Fuji catalog, and the Lick index H$\delta_A$ was computed using pyLick \citep[][]{pylick}. This selection results in 223 post-starburst galaxies. Where objects have duplicate observations with different measured values of EW($\text{[OII]} \lambda3727$), we select the Fuji catalog values corresponding to the dark survey spectra, which have higher SNRs in the $z$ band. This results in the removal of one post-starburst from the \citetalias{soto2025} sample, resulting in a final sample of 222 post-starburst galaxies.

In addition to star formation, $\text{[OII]} \lambda3727$ traces AGN activity. Our cut against $\text{[OII]} \lambda3727$ therefore potentially removes AGN from our sample. We discuss the impact of this cut on our recovered AGN fractions in Section \ref{discussion:selectionmethods}. 

The DESI survey is composed of several sub-surveys, including the luminous red galaxy survey, the emission line galaxy survey, the nearby Bright Galaxy Survey, and a quasar survey \citep[][]{desi_edr}. Objects can be targeted for more than one of these surveys. A total of 160 post-starbursts in the \citetalias{soto2025} sample were included as part of the DESI Luminous Red Galaxy \citep[LRG;][]{zhou2023} sample; 119 were targeted as part of the Bright Galaxy Survey \citep[][]{hahn2023}; 21 were targeted as quasars \citep[][]{chaussidon2023}; 11 were targeted as part of the Emission Line Galaxy survey \citep[ELG;][]{raichoor2023}; 3 were observed as part of DESI commissioning; and 6 galaxies were targeted as part of a secondary DESI survey \citep[see description of secondary surveys in][]{desi_edr}. 7 objects had no targeting information in the DESI EDR catalog. 78 objects in the \citet{soto2025} were previously included in the DESI LRG sample used by \citet{setton2023} to calculate the recently-quenched fraction at $0.4 \leq z \leq 1.3$; we compare to the results of \citet{setton2023} further in Section \ref{discussion:selectionmethods}.

\subsection{Comparison Samples}

Throughout this work, we often compare to a prior sample of low-redshift post-starburst quasars from \citet[]{cales2011}. These objects were selected from the Sloan Digital Sky Survey data release 3 \citep[SDSS DR3;][]{sdss3} quasar catalog as post-starbursts by requiring 1) that the summed equivalent width of the Balmer lines H$\delta$, H$\zeta$, and H$\eta$ be greater than 2 \AA; 2) H$\delta > 1 \textrm{\AA}$; and 3) a Balmer break $>0.9$, where the Balmer break is defined as the flux ratio at rest frame $4035 \textrm{\AA}$ to $3790 \textrm{\AA}$, in addition to further quality cuts described in \citet{cales2011, cales2013}. This selection differs somewhat from ours due to the differing wavelength and redshift ranges covered by the two data sets. We discuss the potential impact of selection effects on our results in Section \ref{discussion:selectionmethods}.

We also  use the Studying Quenching in Intermediate-z Galaxies: Gas, angu$\vec{L}$ar momentum, and Evolution \citep[\squigglecomma;][]{suess2022a} survey as a comparison sample throughout this paper to represent the typical properties of post-starburst galaxies at slightly lower redshift ($0.5 < z \leq 0.9$). This sample was selected based on synthetic UBV colors from SDSS Data Release 14 \citep[][]{sdss14} at $0.50 \leq z \leq 0.94$. We discuss the \squiggle sample selection and its differences from the \citetalias[]{soto2025} sample selection in more detail in Section \ref{discussion:selectionmethods}.

Finally, we compare to the SDSS DR16 quasar catalog \citep{sdss16_qsos} at $0.49 \leq z \leq 1.39$. Throughout this work, we use the spectroscopic properties measured by \citet{wu2022}.

\subsection{Black hole properties}

Throughout this work, we use the measured narrow-line fluxes from the Fuji \texttt{FastSpecFit} value added catalog of the DESI Early Data Release \citep[][J. Moustakas et al. in preparation]{fastspecfit,desi_edr}. We also use the line flux properties from this catalog to select post-starburst quasars (see Section \ref{sec:selection}). However, the Fuji catalog only decomposes the broad and narrow components of the Balmer lines and therefore does not contain separate broad and narrow MgII measurements. Therefore, in order to derive black hole properties and compare these properties with those of the \citet[]{cales2011} sample, we fit broad MgII and \hbeta  line properties for both samples separately using the \texttt{PyQSOFit} quasar SED fitting software \citep[][]{guo2018,shen2019} and \texttt{Astropy} \citep{astropy:2013,astropy:2018,astropy:2022}. 

\texttt{PyQSOFit} takes an input quasar spectrum and decomposes it into quasar components (including continuum, narrow/broad line emission, dust reddening, and FeII complex emission) and a host galaxy component, where one is required for a good fit. We use the \texttt{CZBIN1} quasar templates and the \texttt{PCA} host templates for our fitting; due to the relatively low contribution of the quasar to the spectrum (see further discussion in Section \ref{discussion:selectionmethods}), we exclude the FeII complex and Balmer continuum components. For nine of the ten post-starburst quasars, we are able to decompose the host and quasar components and perform successful fits to the line complexes. For these objects, we fit both broad MgII and broad \hbeta  (where detected) to obtain black hole masses and bolometric luminosities. We include two narrow components for the MgII doublet and one broad component, and we include one narrow and one broad \hbeta  component. We perform MC sampling 200 times per object to obtain the reported $1\sigma$ uncertainties on the host galaxy fraction and line and continuum luminosities. Where no line is detected, we report $2\sigma$ upper limits derived from the total continuum-corrected flux over the 300$\textrm{\AA}$ surrounding the line center minus any detected narrow line flux.

\texttt{PyQSOFit} fails to recover a quasar component in the tenth post-starburst quasar, J0825+8345, likely due to the low quasar fraction in the spectrum (see e.g. Figure \ref{fig:spectra}). We report the $3000\textrm{\AA}$ continuum luminosity and derived Bolometric luminosities for this object as upper limits based on the FastSpecFit values, as they do not decompose the host and quasar components. We report the host fraction for this object as a lower limit based on the lowest recovered host fraction across the other post-starburst quasars in our sample. For all line fits for this object, we use \texttt{Astropy's} \texttt{TRFLSQFitter} function, also with two narrow + one broad component for MgII and one narrow + one broad component for \hbeta.

Throughout this work, we use Equation (10) from \cite{Wang_2009} to obtain the black hole mass from the MgII flux and the continuum luminosity at 3000 \AA. We apply a bolometric correction of 5.15 to $\lambda \text{L}_{3000}$ to compute the bolometric luminosity \citep[]{richards2006}. When using the bolometric luminosity as a probe of accretion rate, we compute the Eddington ratio, $\lambda\equiv \text{L}_{\text{Bol}}/\text{L}_{\text{Edd}}$, where the Eddington luminosity is defined as $\text{L}_{\text{Edd}}\equiv1.51\times10^{38}\text{ (M}_\bullet$/M$_\odot$).

\section{Quasar Selection} \label{sec:selection}

\begin{deluxetable*}{ccccccccc}[t]
    \tablecaption{Post-starburst quasars \label{tab:psb_qsos}}
    \tablehead{\colhead{Name}& \colhead{R.A.}&\colhead{Dec}& \colhead{z} & \colhead{$L_{\text{MgII}}$}& \colhead{$L_{\text{\hbeta, Broad}}$} & \colhead{Log(M$_{\bullet}$/M$_{\odot}$)} & \colhead{Log(L$_{\text{Bol}}/\textrm{ erg s}^{-1}$)} & \colhead{$f_{\textrm{QSO, 4200\AA}}$} \\
     & & & & $\times10^{41}$ erg s$^{-1}$ & $\times10^{41}$ erg s$^{-1}$ &  $-$ & &  }
     \startdata
     \hline
J1107+5244 & 166.79855 & 52.74166 & 0.53 & $15.3_{-0.78}^{+0.8}$ & $3.9_{-0.19}^{+0.19}$ & 7.9 & $44.792\pm0.00119$ & 0.33 \\
J0825+8345 & 126.25802 & 83.76632 & 0.72 & $2.9_{-0.44}^{+0.44}$ & $\leq1.18$ & 7.9 & $\leq44.707$ &  $\leq0.12$\\ 
J1022+3200 & 155.64386 & 32.00912 & 0.75 & $9.0_{-1.0}^{+1.19}$ & $\leq2.73$  & 7.7 & $43.952\pm0.01534$ & 0.12 \\
J0959+0233 & 149.91064 & 2.55465 & 0.75 & $37.0_{-3.48}^{+3.6}$ & $\leq12.21$  & 8.7 & $44.837\pm0.00431$ & 0.32 \\
J1340+3232 & 205.13319 & 32.54882 & 0.80 & $25.5_{-2.29}^{+2.29}$ & $\leq2.53$ &  8.2 & $44.04\pm0.01843$ & 0.13 \\
J1429+0251 & 217.49916 & 2.85215 & 0.82 & $49.3_{-11.46}^{+16.38}$ & $\leq16.27$  & 8.3 & $44.92\pm0.00972$ & 0.18 \\
J0809+3334 & 122.28949 & 33.57154 & 0.85 & $21.2_{-3.64}^{+3.18}$ & $\leq11.05$  & 8.8 & $44.698\pm0.00406$ & 0.27 \\
J1028$-$2447 & 157.05574 & -24.78489 & 0.85 & $71.1_{-3.76}^{+3.26}$ & $\leq8.35$  & 8.4 & $44.548\pm0.00526$ & 0.16 \\
J0548$-$2341 & 87.13614 & -23.68515 & 0.86 & $25.3_{-3.14}^{+6.74}$ & $\leq13.58$  & 8.6 & $44.925\pm0.00284$ & 0.3 \\
J1058+3314 & 164.71118 & 33.24262 & 1.07 & $46.4_{-6.17}^{+6.77}$ & $-$ & 9.0 & $45.329\pm0.00151$ & 0.29 \\
    \enddata
    \tablecomments{Post-starburst quasars selected in this work as described in Section \ref{sec:broadline}. Redshifts are measured from our \texttt{Prospector} fits described in Appendix \ref{appendix:prospector_model}. Reported broad MgII and \hbeta  luminosities and \hbeta-derived black hole masses are computed using the results of our PyQSOFit fitting as described in Section \ref{sec:data}; line luminosity limits are 2$\sigma$ upper limits computed as described in Section \ref{sec:data}. We report an upper limit on the QSO fraction of J0825+8345 as PyQSOFit failed to decompose the host and quasar spectra; we take the minimum QSO fraction across our sample, 0.12, as the minimum recoverable QSO fraction. J1058+3334 has no reported \hbeta  luminosity because the \hbeta  line is not covered by DESI for objects at $z\geq1.02$.}
\end{deluxetable*}

We identify post-starburst quasar hosts using broad line emission indicative of an unobscured AGN. Unobscured AGN emit broad Balmer emission lines from virialized clouds orbiting the central engine \citep[][]{Antonucci1993,urry1995}. \hbeta  is not covered by DESI for the portion of the \citetalias{soto2025} sample at $1.02 \leq z \leq 1.39$, but the MgII$\lambda2796,2803$ doublet is observable over the entire redshift range of that sample. Observations that the width of MgII is often similar to that of \hbeta  \citep[e.g.][]{shen2008} and that its flux changes in response to continuum variability in AGN \citep[though less strongly than e.g. \hbeta, see][]{kokubo2014,sun2015} suggest that MgII is also emitted from the broad line region around AGN, though perhaps not at exactly the same radius as the Balmer lines \citep[see e.g.][]{goad1993,Wang_2009,guo2020}. Broad MgII is therefore commonly used to select quasars in spectroscopic surveys, including in the DESI quasar catalog \citep[][]{alexander2023,chaussidon2023}. We therefore use MgII emission to select quasars in the \citetalias{soto2025} post-starburst catalog.

\begin{figure} 
    \centering
    \includegraphics[width=.9\linewidth]{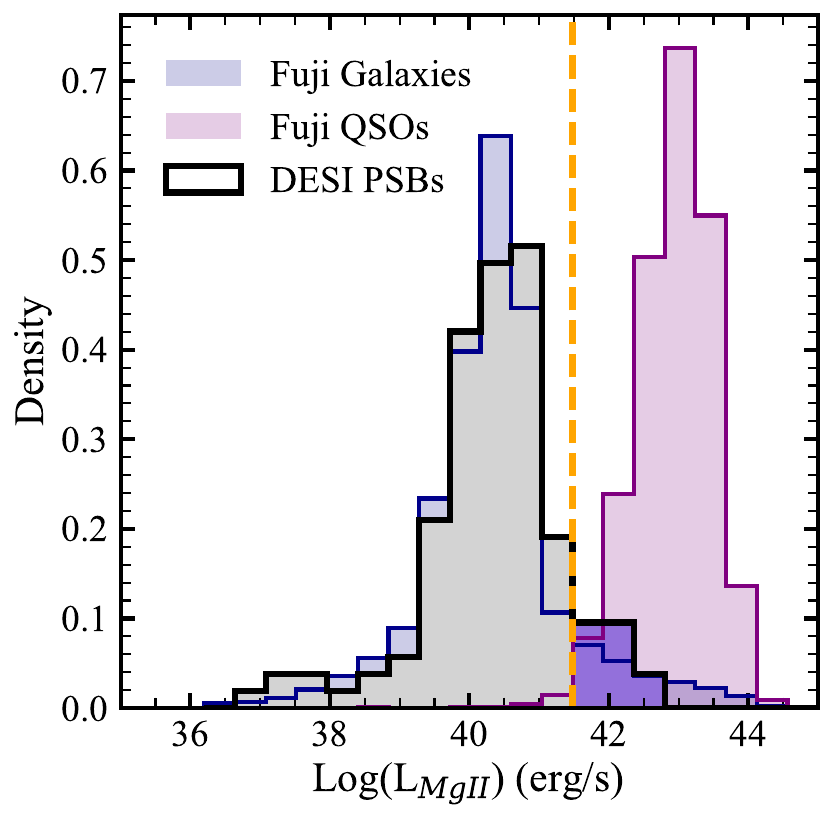}
    \caption{MgII luminosity distribution for DESI galaxies, quasars, and post-starbursts. We select post-starburst quasars by requiring $\textrm{L}_{\textrm{MgII}}>3\times10^{41}\text{ erg s}^{-1}$ and SNR(MgII)$>5$. This selection includes $98.8\%$ (4723/4816) of DESI quasars and $6.23\%$ (267/5804) of DESI galaxies. }
    \label{fig:selection}
\end{figure}

To construct our selection criterion, we test its ability to separate quasars from the DESI Quasar Redshift Catalog \citep[][]{alexander2023,chaussidon2023} from the Fuji catalog galaxies. We begin by removing all quasars from the Fuji galaxy catalog so that we have pure samples of galaxies and quasars. We restrict our analysis to $0.39\leq z\leq1.49$ to match the redshift range of the \citetalias{soto2025} sample. We want to identify the MgII luminosity criterion that maximally separates quasars from galaxies. To do so, we use the F1 score, 

\begin{equation}
    F1\equiv \frac{2\text{TP}}{2\text{TP +FP + FN}}; 
\end{equation}

where a maximal F1 score of 1 indicates maximum balance between the two precision and recall. We construct a set of 30 luminosity thresholds log-spaced between $6.3\times10^{38}$ and $1.0\times10^{44}$ erg s$^{-1}$ and compute the fraction of DESI galaxies and quasars that fall above/below the threshold to compute the true/false positive/negative rates, which we use to compute the F1 score. We find that a MgII luminosity threshold $\textrm{L}_\textrm{MgII} > 3 \times 10^{41}\textrm{ erg s}^{-1}$ maximizes the F1 score with a value of 0.96. To ensure that the measured value of $\textrm{L}_\textrm{MgII}$ does not come from noise in the blue end of the spectrum, we add an additional requirement that SNR(MgII)$>5$. We find that this criterion cleanly separates DESI quasars from DESI galaxies with a completeness of $98.8\%$ and a purity of $93.8\%$ (see Figure \ref{fig:selection}).

We apply this criterion to the \citet[][]{soto2025} catalog and identify ten post-starburst quasars that meet our MgII luminosity threshold. 

\begin{figure*}
    \centering
    \includegraphics[width=\linewidth]{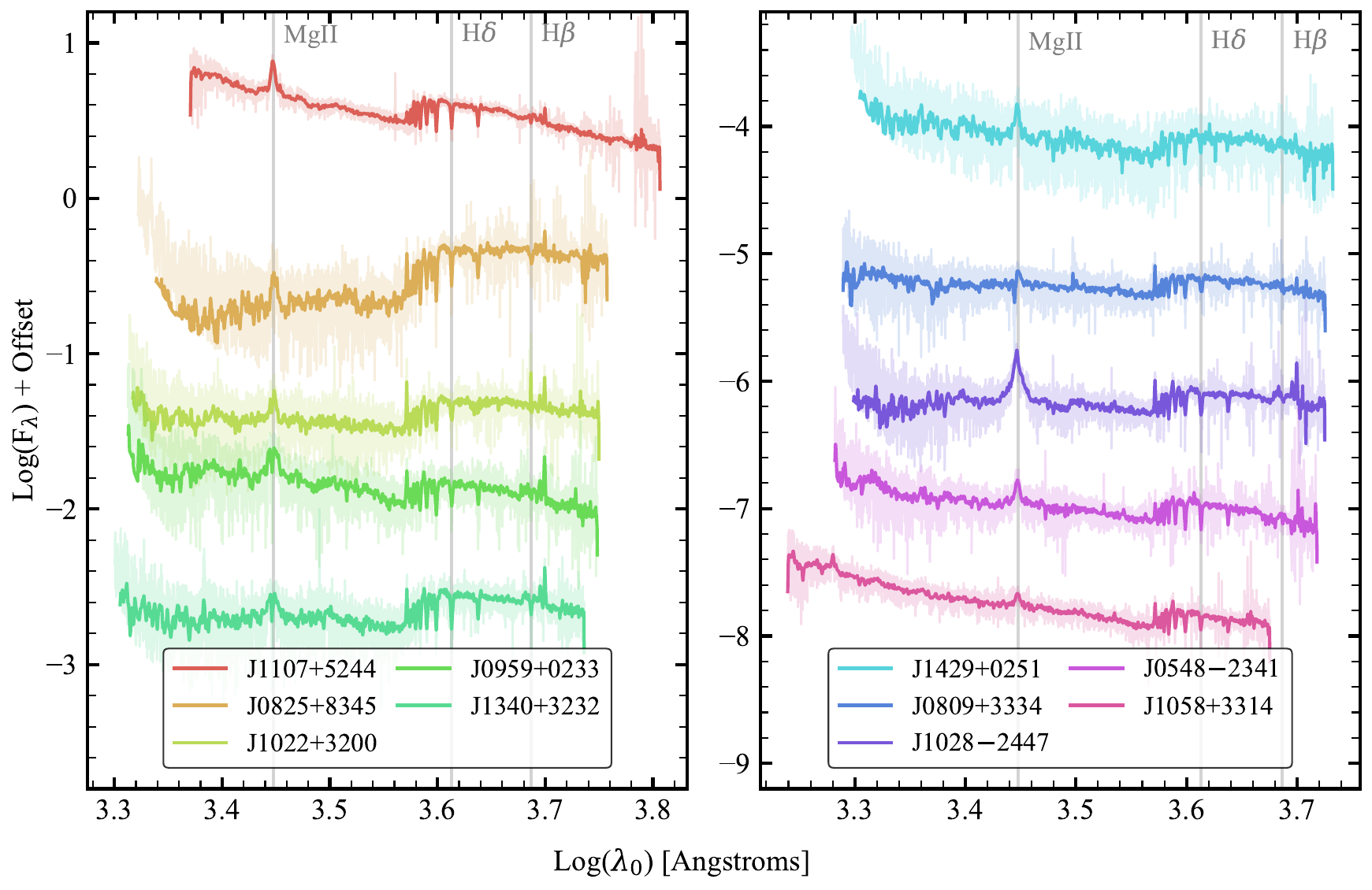}
    \caption{Rest-frame DESI spectra for the post-starburst quasars identified in this work. 5-$\sigma$ Gaussian kernel smoothed spectra are overlaid for ease of feature identification. Spectra are ordered vertically in order of redshift. Post-starburst quasars identified in this study have strong Balmer absorption and lack strong narrow emission lines indicative of star formation; however, they host broad MgII lines that indicate unobscured AGN activity.}
    \label{fig:spectra}
\end{figure*}

The spectra of these post-starburst quasars are shown in Figure \ref{fig:spectra}, and their DESI identifiers, coordinates, redshifts, and MgII luminosities are listed in Table \ref{tab:psb_qsos}. This is the first sample of post-starburst quasars at intermediate redshift. All ten show strong Balmer absorption and Balmer breaks characteristic of a dominant A star population, and lack strong \oii characteristic of star formation, typical of ``E+A" galaxies \citep{dressler1983,Zabludoff1996}. We confirm that these objects are consistent with being Type 1 AGN by measuring the FWHM of MgII using PyQSOFit. Though we did not impose a FWHM requirement as part of our selection, all ten post-starburst quasars have FWHM(MgII)$>2000 \textrm{ km s}^{-1}$, consistent with being unobscured AGN/quasars \citep[e.g.][]{sdssqso}. However, none of the nine post-starburst quasars at $z\leq1.02$ have significant broad Balmer emission in the form of \hbeta  as is normally seen in quasars; we return to this observation in Section \ref{discussion:fading}.

\section{Results} \label{sec:results}

Many methods exist to select AGN in addition to the broad MgII criterion used to select our post-starburst quasar sample. Each of these methods selects a different population of AGN. Broad-line AGN are identifiable only if the AGN is unobscured, meaning the AGN must be oriented such that its own dusty torus does not block the broad line emitting region and that the galaxy nucleus must not be too dusty \citep[][]{Antonucci1993,urry1995,hopkins2005,hickox2018}. Similarly, too much obscuring material can absorb high-energy X-rays, so X-ray selected AGN must either be sufficiently luminous to overcome this obscuration or relatively unobscured \citep[][]{lanz2022}. Mid-infrared and narrow-line selections are less sensitive to obscuring material and therefore make up a larger fraction of AGN \citep[e.g.][]{maia2003,stern2005,lu2010,stern2012}. Roughly 10\% of optically-selected AGN are also detectable in radio wavelengths, due either to a jet or to thermal emission \citep{kellerman1989}. In this section, we apply multiwavelength AGN selections to the \citetalias{soto2025} catalog to determine the AGN occupation fraction in post-starbursts and how it changes based on selection method. A summary table of the fraction of objects identified as AGN in each wavelength regime is given in Table \ref{tab:agnfrac}; classifications for individual objects can be found in Appendix \ref{appendix:agn}. Throughout this work, we report $1\sigma$ binomial errors for all computed AGN occupation fractions.

\begin{deluxetable}{cccccc}[t]
    \tablecaption{AGN properties of post-starbursts} \label{tab:agnfrac}
    \tablehead{\colhead{Selection} & \colhead{$N_{\text{AGN}}$} & \colhead{$N_{\text{Tot}}$} & \colhead{$f_{\text{AGN}}$} } 
     \startdata
     \hline
    MgII & 10 & 222 & $4.5^{+1.6}_{-1.2}\times10^{-2}$  \\
    MEx & 27 & 214 & $1.3^{+0.25}_{-0.21}\times10^{-1}$  \\
    MEx $>$ 3* & 18 & 214 & $8.4^{+2.1}_{-1.7}\times10^{-2}$  \\
    OHNO & 18 & 214 & $8.4^{+2.1}_{-1.7}\times10^{-2}$  \\
    Radio (FIRST)** & 1 (3) & 178 & $5.6^{+9.0}_{-3.5}\times10^{-3}$ \\
    &&& ($1.7^{+1.3}_{-7.3}\times10^{-2}$) \\
    Radio (VLASS)** & 1 (2) & 222 & $4.5^{+7.2}_{-2.8}\times10^{-3}$ \\
    &&& ($9.0^{+8.9}_{-4.5}\times10^{-3}$) \\
    MIR & 6 & 195 & $3.1^{+1.5}_{-1.0}\times10^{-2}$ \\
    X-Ray & 1 & 87 & $1.1^{+1.9}_{-0.7}\times10^{-2}$ \\
    \enddata
    \tablecomments{Multiwavelength AGN fractions for the \citetalias{soto2025} post-starbursts. $N_{\text{AGN}}$ denotes the number of objects classified as AGN in this wavelength regime, while $N_{\text{Tot}}$ denotes the number of objects with sufficient wavelength coverage to determine whether they are AGN. Errors given are $1\sigma$ binomial uncertainties. (*) As discussed in Section \ref{sec:narrowline}, the mass-excitation diagram \citep{juneau2011} predicts all galaxies in our mass range are AGN if they have \oiii/\hbeta$>3$. This criterion is more restrictive than the typical mass-excitation diagram but accounts for uncertainty in recovering galaxy masses in systems with AGN contamination. (**) As discussed in Section \ref{sec:radio}, two of the three post-starbursts detected in radio have radio emission consistent with star formation; however, given the \citetalias{soto2025} selection against star formation, their emission may come at least in part from an AGN. We provide a range of radio AGN fractions to reflect this uncertainty.}
\end{deluxetable}

\subsection{[MgII]$\lambda2803\textrm{\AA}$-Selected Post-Starburst Quasars} \label{sec:broadline}

Based on our MgII luminosity threshold described in Section \ref{sec:data}, we detect quasar emission in \qsorate (10/222) intermediate-redshift post-starburst galaxies from \citet[][]{soto2025}.  

The DESI validation survey was composed of a combination of different targeted galaxy samples, including LRGs, emission line galaxies, the nearby Bright Galaxy Survey, and quasar catalogs \citep[][]{desi_edr}.  We analyze the rate of post-starburst quasars in the \citetalias{soto2025} post-starbursts targeted in each survey; note that individual objects can have more than one targeting classification. We find that $2.5^{+1.6}_{-0.97}\%$ (4/160) of the LRGs, $64^{+13}_{-15}\%$ (7/11) of the ELGs, $0.84^{+1.34}_{- 0.52}\%$ (1/119) of the BGS targets, $43^{+11}_{-10}\%$ (9/21) QSOs, and $29^{+11}_{-8.7}\%$ (6/21) of the secondary targets are post-starburst quasars based on our MgII luminosity threshold. Our AGN occupation fraction is therefore biased by the relative mix of each survey included in the DESI Early Data Release; a large, uniformly selected spectroscopic parent sample would be necessary to infer the fundamental unobscured AGN fraction among post-starbursts.

\subsection{AGN from optical emission line diagnostics} \label{sec:narrowline}

While broad emission lines are an unambiguous signature of an AGN, the broad line region can be obscured by either the AGN's own dusty torus \citep[][]{Antonucci1993,urry1995} or dust inside the host galaxy nucleus \citep[][]{hopkins2005}. Narrow emission line ratios can be used to identify the source of the ionizing continuum as being more consistent with star formation or with AGN activity \citep[][]{baldwin1981}. Most existing narrow-line diagnostics require detection of the \halpha line and \nii doublet, which are too red for the DESI spectrograph at $z>0.49$. We instead use the mass-excitation \citep[MEx;][]{juneau2011} and OHNO \citep{backhaus2022} diagrams, which do not require coverage of \halpha. Throughout this section, we use the FastSpecFit Fuji catalog flux values for each line, using only the narrow component of the \hbeta  line.

The MEx diagram  was developed to identify AGN in intermediate-redshift galaxies where typical diagnostic lines like \halpha and \nii were redshifted out of the optical spectrum. This diagram yields similar results to the BPT diagram for low-redshift galaxies from SDSS \citep[][]{juneau2011} and has been used to distinguish between AGN and star forming galaxies at similar redshift to our sample \citep[e.g.][]{juneau2013,greene2020,lewis2024}. We apply this selection method to the 214 objects in the \citet[]{soto2025} sample at $z \leq 1.02$, as \hbeta  is redshifted out of the observable range above this value. We consider a line to be detected if it has an SNR greater than 3 using the flux and error values provided in the DESI EDR Fuji catalog \citep[][]{desi_edr,fastspecfit}. For an object to be considered an AGN, it must have 1) EW([\textsc{OIII}])$>5$ and 2) be above the MEx line on the \oiii/\hbeta vs. M$_*$ plane \citep[][]{juneau2011}. For the sake of our analysis, we include objects with \hbeta  nondetections and lower limits above the MEx line to be AGN. We find that $13^{+2.4}_{-2.1}\%$ (27/214) of \citet[]{soto2025} post-starbursts at $z\leq1.02$ are considered narrow-line AGN on the MEx diagram. Four of these objects have broad MgII, meaning four of the eight post-starburst quasars with at least one MEx line detected have narrow emission lines consistent with AGN ionization.

\begin{figure*}
    \centering
    \includegraphics[width=.49\linewidth]{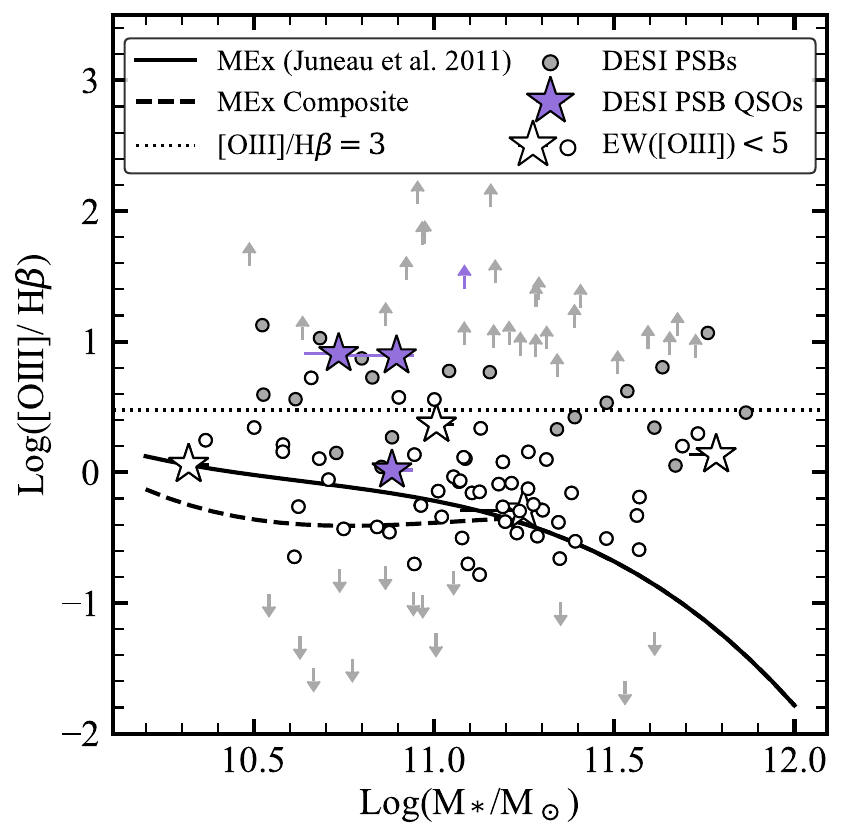}
    \includegraphics[width=.49\linewidth]{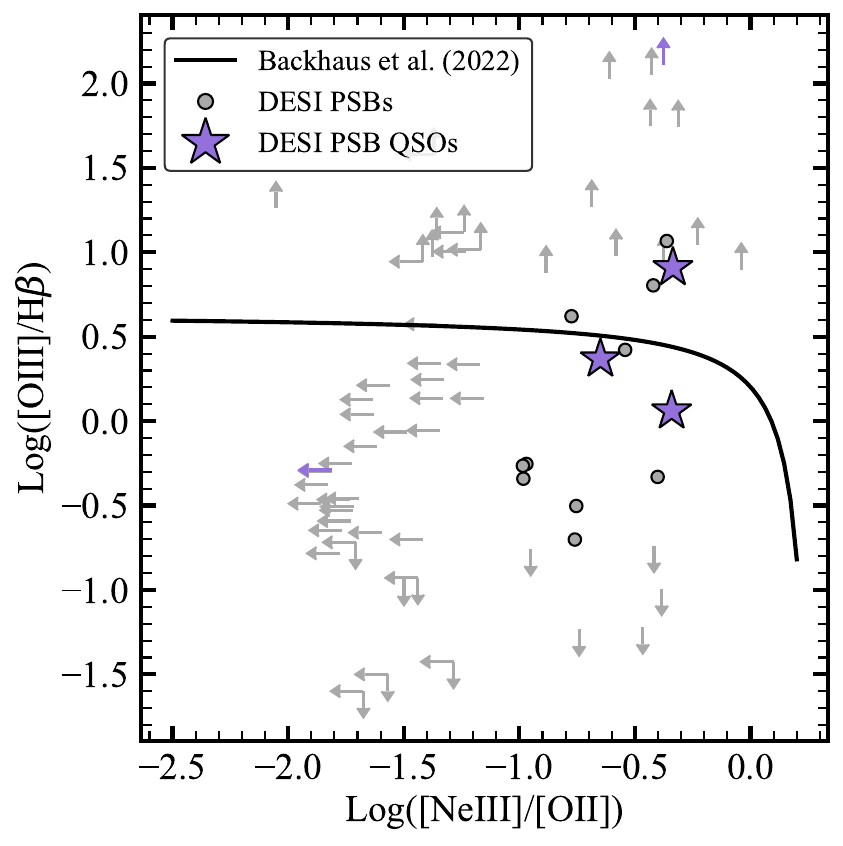}
    
    \caption{(Left) DESI post-starburst galaxies (gray dots) and quasars (purple stars) relative to the MEx diagram introduced in \citet[]{juneau2011}. Limits are designated with arrows; there are no objects with EW(\oiii)$<5$ with limits. Objects above the solid MEx line are considered AGN, with objects between the solid and dashed MEx lines becing consistent with composite and/or LINER emission. $13^{+2.4}_{-2.1}\%$ of the post-starburst sample with detected OIII and \hbeta  are consistent with being narrow-line AGN. Applying the more conservative \oiii/H$\beta > 3$ criterion used in \citet[]{greene2020}, we find that $8.4^{+2.1}_{-1.7}\%$ of \citet[]{soto2025} post-starbursts are narrow-line AGN. (Right) Post-starburst galaxies and quasars on the OHNO diagnostic diagram \citep[][]{backhaus2022}. Objects above the OHNO line are consistent with ionization by AGN. $1.9^{+1.2}_{-0.73}\%$ of post-starbursts have detections in all four lines consistent with being narrow-line AGN; $8.4^{+2.1}_{-1.7}\%$ of the post-starburst sample have either detections or limits that are sufficient to classify them as AGN.}
    \label{fig:linediagnostics}
\end{figure*}

This result depends on the reliability of our stellar mass estimate, and estimating the stellar mass of these objects is complicated by the difficulty of separating the AGN and host contributions to the spectrum when performing SED fitting. For our classifications, we use the inferred stellar mass from a \texttt{Prospector} fit to the data that incorporates the AGN model from \citet[]{bingjiewang2025}; we present the results of this SED fitting in an upcoming work (Verrico et al. in prep.) and briefly describe our SED fitting routine in Appendix \ref{appendix:prospector_model}. The addition of an AGN in stellar population fitting can introduce additional scatter of $\sim0.4$ dex to stellar mass measurements, with no significant offsets in recovered stellar mass for AGN with luminosities up to 90\% of the galaxy luminosity, or with $f_{\text{AGN}}\leq0.47$ \citep[][]{verrico2025}. This is similar to the maximum $f_{\text{AGN}}$ for our sample (see Section \ref{discussion:selectionmethods}); we therefore should not be significantly overestimating the obscured AGN fraction due to overestimates in host galaxy stellar mass caused by AGN contamination. However, for objects with \oiii/H$\beta >3$, galaxies across our mass range (Log(M$_*) \geq 10.25$) are expected to host an AGN \citep[black dotted line in Figure \ref{fig:linediagnostics}][]{juneau2011}; under this more conservative definition, $8.4^{+2.1}_{-1.7}\%$ (18/214) post-starbursts are AGN-like on the MEx diagram, including three of eight post-starburst quasars with at least one line detected.

To avoid relying on a stellar mass measurement, we also use the OHNO diagram to identify narrow-line AGN. The OHNO diagram compares the ratio of \oiii to \hbeta  with the ratio of [NeIII] to OII to determine whether the dominant ISM excitation mechanism is star formation or AGN activity. This diagram takes advantage both of the more extended redshift range for which [NeIII] and OII are observable with optical telescopes ($z\lesssim1.6$ for [NeIII] versus $z\lesssim0.5$ for \nii/\halpha) and the fact that the ratio of [NeIII] to OII should be less affected by dust extinction due to the relatively close wavelengths of the two lines \citep[][]{levesque2014}. We again begin by selecting the 214 objects at $z\leq1.02$. Incorporating objects with upper or lower limits on line fluxes that lead to a definitive AGN or non-AGN classification, we recover an AGN fraction of $8.4^{+2.1}_{-1.7}\%$ (18/214), though objects without detections of all four lines are unlikely to be strong AGN (see discussion in Section \ref{discussion:fading}). One of the nine post-starburst quasars with coverage of the \hbeta  line has OHNO line ratios consistent with AGN-dominated ionization. If we include only objects with strong detections in all four lines, we find that $1.9^{+1.2}_{-0.73}\%$ (4/214) of the \citetalias{soto2025} objects are AGN-like on the OHNO diagram.

Finally, we use the \nev$\lambda3427$ coronal line to probe AGN activity in our post-starburst quasar sample. This line is used as an AGN probe in massive galaxies \citep[e.g.][]{gilli2010,cleri2023,euclid2025,trakhtenbrot2025,barchiesi2026}, including in high-redshift studies of green valley and/or post-starburst galaxies \citep[e.g.][]{vergani2018,valentino2026}. We consider an object to have a \nev detection if the line has an SNR$>3$. We find that $4.5^{+1.6}_{-1.2}\%$ (10/222) of the objects in the \citetalias{soto2025} sample have strong \nev detections, including $30^{+16}_{-12}\%$ (3/10) of post-starburst quasars.

\subsection{Radio activity and variability} \label{sec:radio}

Three objects in the \citet[]{soto2025} sample are detected in the FIRST and VLASS radio surveys. Three of the 178 post-starbursts covered by the FIRST footprint are detected (J1058+3314, J1409-0139, and J1611+4415) and two of the 222 post-starbursts covered by the VLASS footprint are detected in both surveys (J1058+3314 and J1611+4415); the third object, detected in FIRST, J1409-0139, is covered by the VLASS footprint but is not detected in VLASS. This results in a $1.7^{+1.3}_{-0.73}\%$ detection rate in FIRST and a $0.90^{+0.89}_{-0.45}\%$ detection rate in VLASS, somewhat lower than the $\sim4\%$ FIRST detection rate in intermediate-redshift post-starbursts in \citet[]{greene2020} and similar to the VLA detection rate in $0.5 < z < 3.0$ post-starbursts in \citet[]{patil2026} ($0.8\%$, and see Figure \ref{fig:luminosity_limits}). One of the objects detected in FIRST and VLASS is the post-starburst quasar J1058+3314; the other two objects lack broad MgII emission. J1409-0139 is also classified as an AGN based on its location on the MEx diagram, though it has \oiii/\hbeta$<3$ and is therefore likely a low-luminosity AGN and/or LINER. J1611+4415 is not classified as an AGN based on any of the other criteria used in this work. 

Some optically-selected post-starburst galaxies may still retain significant dust-obscured star formation after a significant merger, though the exact role of dust in the post-starburst phase is debated \citep{smercina2018,baron2023,setton2025,hewitt2026}. Dust-obscured star formation can contribute to the radio luminosity of the galaxy, meaning some of the radio luminosity of our sample may come from star formation rather than AGN \citep[][]{patil2026}. Radio emission from star formation can also persist for at least $\sim150$ Myr, meaning an excess of radio relative to instantaneous SFR in post-starbursts may be due to the recent burst \citep{arangotoro2023}. We test whether the radio emission is consistent with star formation or AGN in each of these three sources to find the rate of radio AGN in the \citetalias{soto2025} sample.

First, we employ the use of radio spectral slopes to determine whether these objects are consistent with being radio AGN. We measure the power law slope $\alpha$ for each object using the FIRST data at 1.4 GHz and the VLASS data at 3 GHz. As mentioned above, J1409-0139 lacks a detection in VLASS. Some low-redshift post-starbursts have been found to host variable radio sources, which could be either from tidal disruption events or newborn AGN \citep[][]{French2025}. Objects that are detected in VLASS but not FIRST are generally considered to be radio variable, but objects that are detected in FIRST but not VLASS might instead have steep spectral slopes \citep{nyland2020}. We test whether the object detected in FIRST but not VLASS is consistent with having a steep slope, we use the median 3$\sigma$ upper limit from VLASS epochs 1--4 to compute a spectral slope $\alpha\leq-2.26$ for this source. This slope is consistent with a starburst; see \citet[]{bondi2007}. We therefore do not require variability to explain the observed radio properties of the objects in the \citetalias{soto2025} sample. We find $\alpha=0.4\pm0.2$ for J1611+4415 and $\alpha=-0.1\pm0.1$ for J1058+3314. These slopes are both higher than the canonical $\alpha=-0.7$ value for star forming galaxies, suggesting an AGN contribution to the radio spectrum. However, there is significant overlap between AGN and star-forming galaxy spectral slopes at these values \citep{condon1992}; we therefore must turn to star formation rate diagnostics to determine the source of the radio emission.

 For one of the three detected objects, we have already determined that star formation is the likely culprit for its radio emission. To test whether the observed radio emission in the other two objects could be consistent with star formation, we compute the predicted star formation rate from $L_{1.4 \text{GHz}}$ following the methods of \citet[]{patil2026}; in brief, we combine the $L_{\text{IR}}-$SFR correlation from \citet{kennicutt1998} with the IR-radio correlation for star-forming galaxies \citep[e.g.]{helou1985,condon1992,yun2001} to compute the expected SFR if all radio emission comes from star formation in J1058+3314 and J1611+4415. As we note above, past star formation can contribute to the observed radio luminosity of galaxies with declining SFRs; we therefore assess only whether the radio-derived SFRs are physically plausible for massive galaxies at intermediate redshift, not whether they are consistent with being post-starburst. We obtain star formation rates of $\sim10^{4.4}$ M$_\odot$ yr$^{-1}$ and $\sim10^{2.8}$ M$_\odot$ yr$^{-1}$ for J1058+3314 and J1611+4415, respectively. The former is three orders of magnitude above the star-forming main sequence, while the latter is consistent with a main-sequence galaxy at this redshift \citep{Whitaker2012b}. Given the high star formation rate necessary to reproduce the radio emission in J1058+3314 and the fact that it hosts an active quasar, we conclude AGN activity is very likely the dominant source of radio emission in this object. However, J1611+4415 may have either contribution from an AGN or past and/or dust-obscured star formation, or some combination of the two. This results in a recovered radio AGN fraction of 1/176, or $0.57^{+0.91}_{-0.35}\%$. A more detailed stellar population analysis that accounts for the relative contribution of the AGN and the galaxy to the optical spectrum would be necessary to determine how much radio emission comes from the AGN; we present detailed stellar population fits to these objects in an upcoming work (Verrico et al. in prep). 

\subsection{Mid-IR AGN} \label{sec:mir}

AGN have extremely red MIR colors out to $z\sim3.5$ due to the slope of the dusty torus SED; because of this, the Wide-field Infrared Survey Explorer \citep[WISE;][]{wright2010} W1 band at 3.4 microns and W2 band at 4.6 microns have been used to identify AGN. A WISE color cut of W1-W2 $\geq0.8$ selects luminous AGN regardless of obscuration with purity of $95\%$, though it can miss a significant number of lower-luminosity AGN in the redshift range $0.5 < z < 1.5$ \citep[][]{stern2012}. A magnitude-dependent criterion like those presented in \citet{assef2018} can more reliably recover some lower-luminosity AGN, though \citet{assef2013} found no change in completeness with W2 magnitude. Therefore, the fractions of MIR-luminous AGN reported here should be taken as a lower limit, as less luminous AGN in host-dominated systems as well as AGN in dust-poor systems will not be selected.

196 of the 222 objects in the \citet[]{soto2025} catalog are included in the ALLWISE source catalog \citep[][]{allwise}, including 8 of 10 post-starburst quasars. Of these, 192 are detected in W1 and W2 with SNR$\geq3$, including all 8 post-starburst quasars with WISE coverage.

$3.1^{+1.5}_{-1.0}$ (6/192) of the WISE-detected post-starbursts are classified as AGN based on the \citet{stern2012} MIR color selection. Two additional post-starbursts are selected by the \citet{assef2018} R75 criterion, for a recovered MIR AGN fraction of $4.2^{+1.7}_{-1.2}\%$. None of the detected post-starburst quasars are classified as WISE AGN based on their MIR colors. Considering again that this fraction includes only AGN-dominated and/or dusty systems, the true fraction of dust-obscured AGN in this sample is likely higher. Detailed SED modeling would therefore be necessary to fully characterize the AGN properties of this sample.

\subsection{X-Ray} \label{sec:xray}

One post-starburst quasar (J0959+0233) is marginally detected in eROSITA ERASS Data Release 2 \citep{erositadr2} in the 0.2-2.3 keV band with a detection likelihood of 11.3 of the 87 objects within the eRASS footprint. This corresponds to a detection rate $1.2^{+1.8}_{-0.71}\%$ (1/87) among post-starburst galaxies and $14^{+18}_{-8.7}\%$ (1/7) among the MgII-emitting post-starbursts in the eROSITA footprint. The detected post-starburst quasar does not have enough detected photons for a spectrum and has an X-ray luminosity of $(8\pm3)\times10^{43} \textrm{ erg s}^{-1}$. Ten of the objects with upper limits are flagged for having a close neighbor.

\section{Discussion}
\subsection{Multiwavelength AGN Fractions} \label{discussion:fraction}

\begin{figure*}
    \centering
    \includegraphics[width=\linewidth]{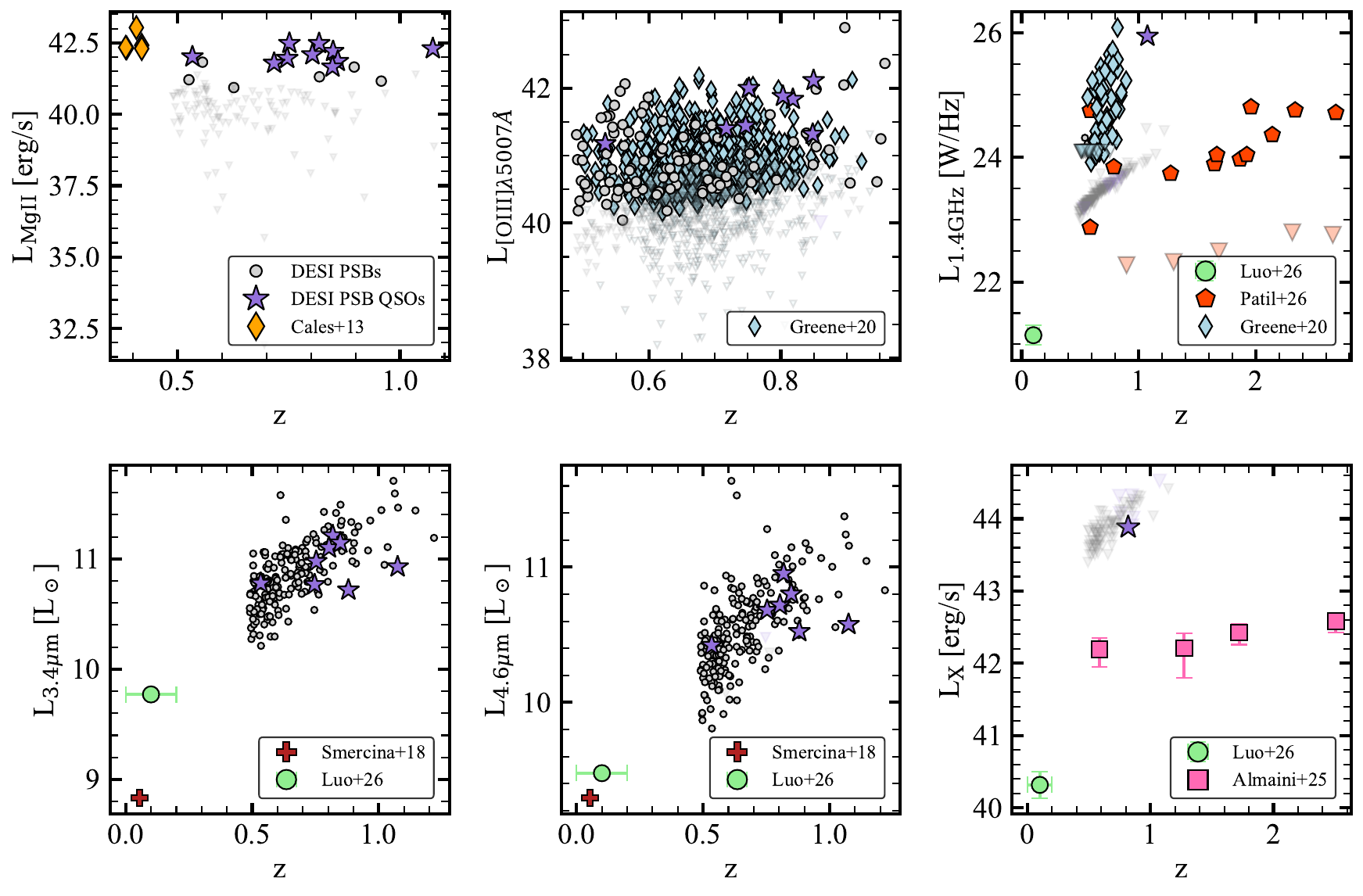}
    \caption{Luminosities of the AGN selected by different criteria in the \citet[]{soto2025} sample. Post-starburst galaxies from \citet[]{soto2025} are gray circles; post-starburst quasars are purple stars. From top right to bottom left, we show the luminosity of broad MgII, narrow \oiii, FIRST data at 1.4 GHz, WISE 1 and WISE 2, and eROSITA soft X-ray (0.2--2.3 keV) as a function of redshift. Upper limits are shown as down-facing triangles. For comparison, we show literature values for the luminosities in these bins for post-starburst galaxies ranging from $0\lesssim z\lesssim3$ from \citet{smercina2018,greene2020,almaini2025,luo2026} and \citet{patil2026} as well as the low-redshift post-starburst quasars from \citet{cales2011}. We discuss differences in selection methods and recovered AGN fractions for these samples in Section \ref{discussion:selectionmethods}.}
    \label{fig:luminosity_limits}
\end{figure*}

Based on the multiwavelength AGN criteria used in Section \ref{sec:results}, a total of 57 unique post-starbursts from the \citetalias{soto2025} catalog are selected based on these criteria. This results in a total AGN rate of $26^{+3.0}_{-2.8}\%$.

Throughout this work, we compare to AGN fractions of post-starburst samples at lower redshift. Comparing multiwavelength AGN occupation fractions is challenging, as changes in obscuration, AGN luminosity, host galaxy contribution, and instrument sensitivity can strongly bias the measured occupation fraction at different parts of the SED. For our purposes, X-ray and broad-line signatures of AGN likely point to the unobscured AGN fraction, while MIR- and narrow-line-selected AGN trace the entire AGN population, including obscured AGN. Different post-starburst selection methods can also bias the AGN occupation fraction, as some post-starburst selection methods allow for more emission from AGN \citep[e.g.][]{Alatalo2016a,Starecheski2026}. It is therefore challenging to determine whether the AGN fraction of post-starburst galaxies has changed with cosmic time, and therefore whether and when AGN feedback has played an important role in rapid quenching. Here, we compare our measured AGN fraction to results from previous works; we show the luminosity and redshift distribution of our sample in the context of other post-starburst AGN surveys in Figure \ref{fig:luminosity_limits}.

\subsubsection{Unobscured AGN} \label{discussion:unobscuredagn}

Based on their broad-line and X-ray properties, we find that 10 post-starbursts in the \citetalias{soto2025} catalog are consistent with being unobscured AGN, leading to a recovered unobscured AGN fraction of \qsorate. In the local Universe, $\sim5-15\%$ of post-starbursts have been found to host AGN with $L_{X} \gtrsim 10^{40}\text{ erg s}^{-1}$ based on detections in Chandra (0.5-7 keV) and eROSITA \citep[0.2-2.3 keV;][]{georgakakis2008,luo2026}. Using deep Chandra observations at 0.5-7 keV, \citet{brown2009} found that K+A galaxies were more likely to host X-ray sources than other galaxies, with as many as a third of $M_r < -22$ K+A galaxies having X-ray luminosities $L_X \sim 10^{42} \text{ erg s}^{-1}$; \citet{lanz2022} found that as many of half of local post-starbursts are consistent with hosting low-luminosity ($L_{2-10 \text{ keV}} < 10^{42}\text{ erg s}^{-1}$) AGN. Low-redshift stacking results are also consistent with a population of AGN in post-starbursts, despite low detection fractions \citep{georgakakis2008,luo2026}. These results suggest a minority of low-z post-starbursts host X-ray luminous AGN, though many may host lower-luminosity and/or obscured AGN. Out to cosmic noon, X-ray AGN again occur in a few percent of post-starbursts \citep[e.g.][]{almaini2025}, though these rates are limited by the sensitivity of X-ray telescopes. Stacking has been employed to find average X-ray luminosities consistent with low-luminosity or obscured AGN in post-starbursts at all redshifts \citep[][]{georgakakis2008,almaini2025}, though \citet[][]{luo2026} found stacked X-ray luminosities that could instead be explained by obscured star formation. These studies are consistent with our post-starburst quasar fraction (\qsorate) and somewhat higher than our X-ray detection rate ($1.2^{+1.8}_{-0.7}\%$), though we are limited by the lack of X-ray coverage by more sensitive X-ray telescopes like Chandra and NuSTAR. In Figure \ref{fig:luminosity_limits}, we show the X-ray luminosities and luminosity limits for objects in the \citet[]{soto2025} catalog. These upper limits are not very constraining, as they correspond to much higher X-ray luminosities than those observed in post-starburst galaxies at this redshift range \citep[][]{almaini2025}. Without more sensitive X-ray data, we cannot confirm or rule out the presence of lower luminosity or obscured X-ray AGN consistent with the stacking results of \citet{almaini2025} and \citet{luo2026} in our sample. However, the low rate of X-ray luminous post-starbursts in the \citetalias{soto2025} sample implies that X-ray bright AGN are rare in the post-starburst phase and may indicate that the AGN is short-lived relative to the $10^{8-9}$-year post-starburst phase.

Broad-line post-starburst quasars in the literature were selected from quasar catalogs rather than from post-starburst catalogs, making it difficult to directly compare the post-starburst quasar fraction across redshifts; however, post-starburst galaxies were found to make up $\sim3\%$ of quasar hosts at $z<0.4$ \citep{cales2013}, which is 10x higher than the $\sim0.2\%$ of SDSS galaxies that are classified as post-starburst \citep[][]{french2016}. Recently, \citet[]{krishna2025} and \citet{sun2026} found that as many as a quarter of quasar hosts at $z\lesssim1$ were post-starburst. We cannot compare the quasar rate in intermediate-redshift post-starbursts, as previous surveys select against QSO emission, which we discuss in Section \ref{discussion:selectionmethods}. Future surveys that uniformly select post-starburst quasars and galaxies will be necessary to compute the real broad-line AGN fraction in post-starburst galaxies at all redshifts.

\subsubsection{Narrow-line and MIR AGN} \label{discussion:obscuredagn}

Estimates of the AGN fraction in post-starbursts from narrow emission lines vary between studies, with some studies finding as many as half of post-starbursts may host Seyferts or LINERs based on their emission line ratios \citep[][]{brown2009, pawlik2018}. While narrow emission lines may trace obscured AGN, AGN-like emission ratios can also persist after the central engine has turned off \citep[e.g.][]{bennert2006,greene2011} or be caused by shocks \citep[e.g.][]{Alatalo2016a} or evolved stellar populations \citep[][]{cidfernandes2011}. Indeed, studies with more stringent cuts against LINERs or non-nuclear emission find lower AGN fractions of $\sim4-5\%$ in local post-starbursts \citep[e.g.][]{french2023,Starecheski2026,huang2026}. At intermediate redshift, \citet[]{greene2020} found that $5.3\pm0.7\%$ of color-selected post-starbursts hosted a narrow-line AGN, with a higher AGN fraction in objects with lower D$_{n}4000$ values which traces younger post-burst age. \citet{valentino2026} found a similar preference for AGN in post-starbursts with lower D$_{n}4000$ at $1.5 \leq z \leq 4$, where $\sim7\%$ of quiescent galaxies were detected in \nev, while \citet[]{skarbinski2026} found about half of post-starbursts at $1 < z < 3$ have narrow line ratios consistent with AGN.

MIR colors consistent with AGN torus emission have also been used to identify AGN in post-starbursts. \citet[]{meusinger2017}, \citet[]{smercina2018}, \citet[]{french2023} and \citet[]{luo2026} found $\sim3-5\%$ of post-starbursts at $z<0.4$ hosted MIR-selected AGN, suggesting low-redshift post-starbursts have a low AGN fraction. \citet[]{alatalo2017} and \citet[]{meusinger2017} also found that the mean MIR SEDs of post-starbursts were consistent with widespread low-level AGN activity. At higher redshift, \citet[]{hewitt2026} found that $3\%$ (1/33) of massive post-starbursts at $0.5 < z < 3.0$ had MIR excesses consistent with AGN. 

We find that $\sim8-13\%$ of post-starbursts in the \citet[]{soto2025} catalog are narrow-line AGN and $\sim3\%$ of post-starbursts are classified as AGN based on their WISE colors. These fractions are slightly higher than the \citet[]{greene2020} narrow-line AGN fraction for intermediate-redshift post-starbursts and are roughly consistent with many results at low and high redshift. We likely find a lower narrow-line AGN fraction than the $\sim50\%$ found at low and high redshift using the BPT and WHAN diagrams because we require EW(\oiii)$>5$, a requirement used in the \citet[]{greene2020} but not in the BPT and WHAN diagrams used to select narrow-line AGN at Cosmic Noon by \citet[]{skarbinski2026}. This requirement, as well as the more stringent MEx definition used by \citet[]{greene2020}, likely selects Seyfert AGN rather than LINERS; indeed, objects with  EW(\oiii)$\leq5$ in Figure \ref{fig:linediagnostics} (shown with white circles) cluster near the MEx diagnostic line where LINERs are more likely to be found \citep[][]{juneau2011}.When we remove this requirement, we find a significantly higher AGN fraction of $50\pm3.4\%$ (107/214), consistent with the results of \citet{skarbinski2026}. However, due to a lack of coverage of [NII], [SII], and \halpha, we are unable to directly compare our narrow-line AGN fraction with fractions that include or exclude lower-luminosity AGN and LINERs.

\subsection{Post-starburst quasars are missing from most surveys} \label{discussion:selectionmethods}

Post-starburst selection methods can bias the measured AGN fraction during the post-starburst phase. Traditional E+A selection methods at low redshift use either [OII] or \halpha to trace star formation \citep[e.g.][]{Zabludoff1996,Goto2005}, meaning many E+A-selected post-starburst samples select against AGN activity (including the \citetalias{soto2025} catalog). Similarly, selection with color-color diagrams \citep[e.g.][]{kriek2010,whitaker2012a,belli2019,suess2022a} removes objects with blue continuum that may originate from star formation or AGN emission. However, selection methods that allow for increased emission from shocks/AGN have been developed \citep[e.g.][]{Alatalo2016a}, and the Principal Component Analysis (PCA) technique used to select post-starbursts in e.g. \citet[]{Wild2007,wild2009,rowlands2015} and the PCA-derived supercolor selections should also allow for galaxies with more blue continuum from AGN or from small amounts of ongoing star formation. The supercolor selection technique developed in \citet{wild2014} and used in \citet{skarbinski2026,hewitt2026} also allows for a small amount of AGN contribution to the optical spectrum, though highly-accreting ($\gtrsim10\%$ L$_{\text{Edd}}$) and/or relatively unobscured (A$_\nu < 2$) will be excluded \citep{almaini2025}. This further complicates comparisons between our recovered multiwavelength AGN fractions and other literature on the topic.

\begin{figure}
    \centering
    \includegraphics[width=\linewidth]{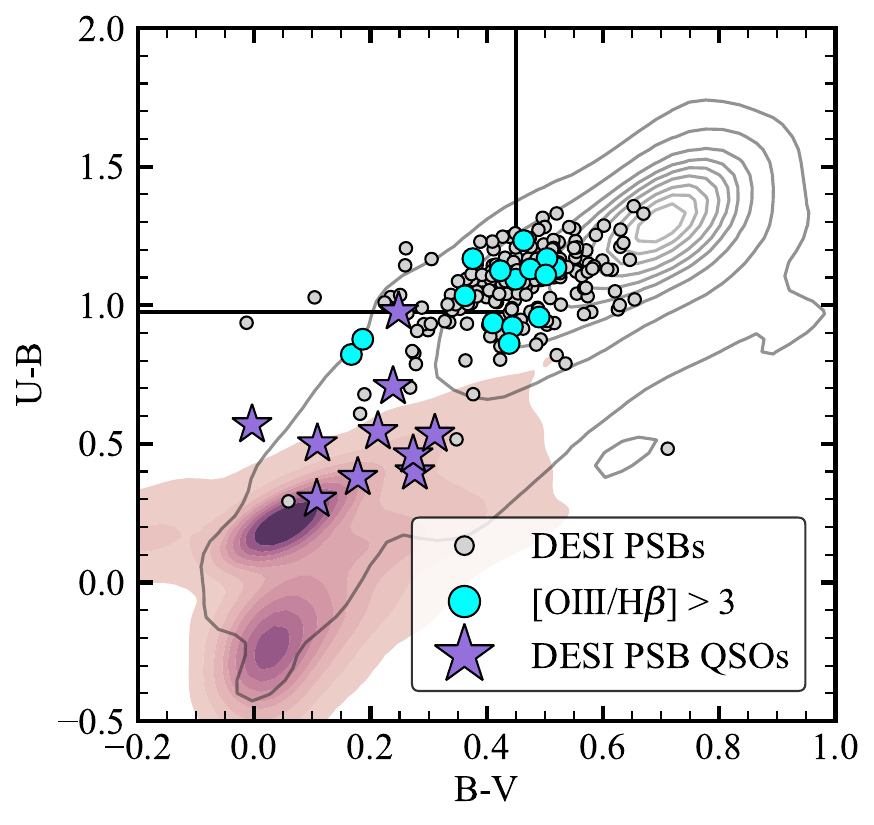}
    \caption{UBV color-color diagram from \cite{kriek2010} used to construct the \squiggle intermediate-redshift post-starburst sample \citep[][]{suess2022a}. The DESI EDR galaxy sample is shown in gray contours, and the DESI EDR quasar sample is shown in purple contours \citep[][]{desi_edr}. Colors are computed using synthetic rest-frame Johnson's UBV magnitudes from the Fuji value-added catalog \citep[][J. Moustakas et al. in preparation]{fastspecfit}. Post-starburst galaxies hosting narrow-line AGN (\oiii/\hbeta$>3$, EW(\oiii)$>5$) are shown in blue. All but one of the detected post-starburst quasars fall outside of the UBV selection used in \citet[]{suess2022a} (upper left box), explaining why they have not previously been identified at this epoch.}
    \label{fig:colorcolor}
\end{figure}

To quantify this effect, we first reproduce the color selection used for the $z\sim0.7$ \squiggle post-starbursts used in \citet{greene2020}. This sample was selected using synthetic UBV filters first developed by \citet[]{kriek2010} designed to capture objects with large Balmer jump and a flat slope of the spectrum redward of the Balmer break \citep[$U_m-B_m>0.975$ and $B_m-V_m<0.45$;][]{suess2022a}. To test how our AGN fraction would have changed if we instead used color-color selection, we use the synthetic Johnson's UBV magnitudes provided in the Fuji VAC \citep[][J. Moustakas et al. in preparation]{fastspecfit,desi_edr}. These filters are similar to, though not identical to, the synthetic filters constructed by \citet[]{kriek2010}. 69 of the 222 \citet[]{soto2025} post-starbursts are selected based on this color cut, or $30\%$. In Figure \ref{fig:colorcolor}, we show that all but one of the post-starburst quasars identified in this work would have been excluded from the \squiggle post-starburst sample \citep[][]{suess2022a}, leading to a broad-line AGN rate of $1.5^{+2.3}_{-0.91}\%$. Using the same MEx classification used in \citet[]{greene2020} on the \squiggle sample, we find 4/68, or $5.9^{+3.5}_{-2.4}\%$, of UBV-selected \citet[]{soto2025} post-starbursts are selected as narrow-line AGN, consistent with both our narrow-line AGN fraction and that measured in \citet[]{greene2020}. Overall, it does not appear there is a significant bias in the \squiggle selection against selecting narrow-line AGN, but unobscured AGN/quasars would be almost entirely excluded from their sample.

Post-starbursts can also be selected based on SED fitting to identify objects that recently formed a large fraction of their mass and then quenched \citep[e.g.][]{setton2023,shepherd2026a,Starecheski2026}. Accurately recovering the recent star formation of AGN host galaxies can be challenging due to the similarity between the spectral features of AGN and active star formation; broad-line AGN can introduce additional uncertainty \citep[e.g.][]{verrico2025} or offsets \citep[e.g.][]{cardoso2017} into recovered galaxy masses, SFRs, and star formation histories. We use the DESI post-starbursts identified by \citet{setton2023} to determine whether objects selected based on their derived star formation histories have a different quasar fraction than the spectroscopically-selected \citetalias{soto2025} sample. \citet{setton2023} used the DESI survey validation LRG sample at $0.4 \leq z \leq 1.3$ as a parent sample. They fit the 17,217 survey validation LRGs with \texttt{Prospector} stellar population synthesis software \citep{johnson17,leja17,johnson2021} using the flexible-bin star formation history template described in \citet{suess2022b}. To deal with potential contribution from narrow-line AGN, they marginalized over nebular emission, but they did allow for any UV-to-optical contribution from AGN in their modeling. They performed several post-starburst selections; for the purpose of our comparison, we focus on their fiducial selection which required 1) that the galaxy be quiescent by restricting their study to objects at least 0.6 dex below the \citet{leja2022} star-forming main sequence and 2) that the object have formed at least 10\% of their mass in the last Gyr before observation. To identify unobscured AGN in this sample, we identify post-starbursts in the \citet{setton2023} sample with $\textrm{L}_{\text{MgII}}>3\times10^{41} \textrm{ erg s}^{-1}$. We find 12,573 quiescent galaxies and 695 post-starburst galaxies of the 17,157 galaxies with successful \texttt{Prospector} fits in the \citet{setton2023}. We do not identify any post-starburst quasars among this post-starburst sample, despite the fact that three of the DESI post-starburst quasars in our sample (J0959+0233, J0809+3334, and J1028$-$2447) are in the \citet{setton2023} catalog. These three objects were classified as star-forming based by \citet{setton2023}, likely due to the lack of AGN model in their fits. This results in a significantly lower recovered limit on the AGN fraction of $\leq0.14\%$ for this stellar population synthesis-based catalog.

\begin{figure}
    \centering
    \includegraphics[width=\linewidth]{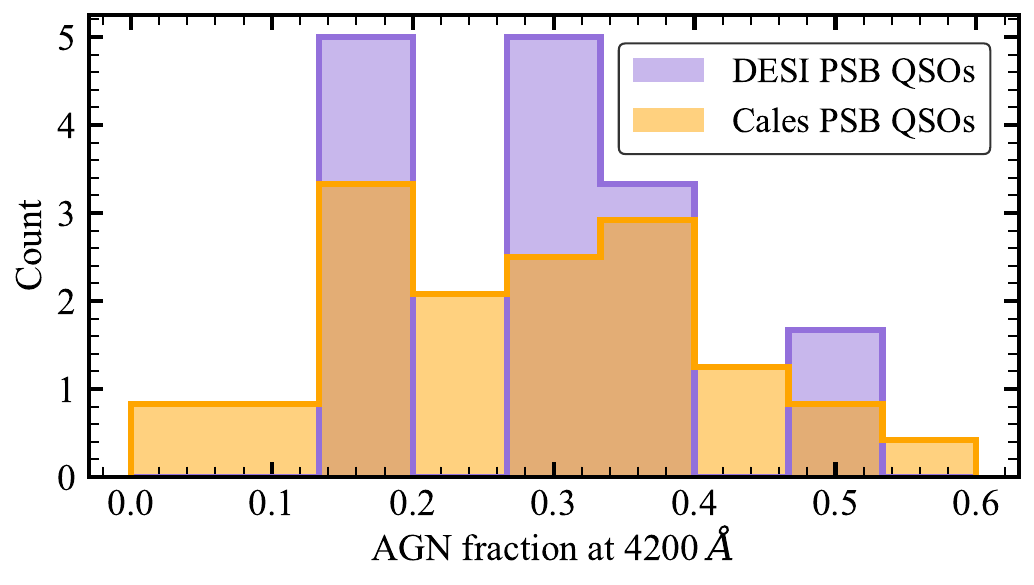}
    \caption{The AGN fraction of the total integrated light measured using PyQSOFit at $4200\textrm{\AA}$ for our post-starburst quasar sample (purple) and the \citet[]{cales2013} low-redshift post-starburst quasars. We find that H$\delta_A$ selection methods no longer select post-starbursts when the AGN fraction is greater than $50\%$, consistent with the AGN fraction of our sample.}
    \label{fig:agnfraction}
\end{figure}

But what about spectroscopic selection? Here, additional OII and/or \halpha emission from the AGN can mimic star formation, while infilling of H$\delta$ can mask the signatures of a dominant A-star population. To test this possibility, we generate two single stellar populations (SSPs) using \texttt{Prospector} with an age of 300 Myr and masses of $10^{10}$ and $10^{12} \text{ M}_\odot$. To these, we add the quasar template presented in \citet[]{shen2011}, normalized to AGN fractions of [0.75, 0.66, 0.50, 0.40, 0.33, 0.25, 0.10] (measured at $4200\text{\AA}$). We then test at what AGN fraction the value of H$\delta_{\text{A}}$ drops below 4. We find that post-starbursts are identifiable as long as the AGN emits $\lesssim50\%$ of the total integrated luminosity at $4200\text{\AA}$ for both SSPs, corresponding to a bolometric luminosity range of $44.6 \lesssim \textrm{Log(L}_{\text{Bol}}/\textrm{erg s}^{-1}) \lesssim 46.6$ across the mass range $10 <\textrm{Log(M}_{*}/\textrm{M}_{\odot}) < 12$. From our PyQSOFit results, we find that the post-starburst quasars in our sample emit between $15\%$ and $51\%$ of the total integrated light at $4200\text{\AA}$ (Figure \ref{fig:agnfraction}), while the \citet[][]{cales2013} post-starburst quasars emit between $0.4\%$ and $59\%$ of the total integrated light at $4200\text{\AA}$ (note that the \citet{cales2011} objects were selected using a cut on all Balmer lines rather than just H$\delta$, which may explain why they can have more AGN contribution). The \citetalias{soto2025} sample, as well as other H$\delta$-selected samples, is therefore missing post-starbursts that are quasar dominated.  

The \citet{shen2011} quasar template does not have significant \oii emission, so we cannot use it to determine how our cut against \oii biases our recovered AGN fraction. As many as half of H$\alpha + \text{K}+\text{A}$-selected post-starbursts have significant \oii emission, with the majority of \oii emission classified as AGN- or LINER-like on a BPT diagram \citep{yan2006}. On the other hand, recent work by \citet{wu2026} employs kinematic decomposition to separate AGN and star-formation dominated \oii emission and finds \oii emission in quasar hosts is dominated by star formation. To test how our \oii selection biases our recovered quasar fraction, we estimate the median and maximum \oii luminosities of quasars in the Fuji catalog with EW(\oii)$\leq3$, the same cut used to select our post-starburst sample. We find that Fuji quasars with EW(\oii)$\leq3$ have a maximum L(\oii)$=6.6\times10^{42} \textrm{ erg s}^{-1}$. There are no standard bolometric corrections for \oii$\lambda3727\textrm{\AA}$ due to the relatively large expected contribution from star formation in normal AGN \citep[e.g.][]{zhuang2019,wu2026}; however, assuming all of the observed \oii emission comes from AGN in the Fuji quasars, we can predict the amount of expected \oiii$\lambda 5007\textrm{\AA}$ emission following the predicted NLR \oii/\oiii ratio from \citet{zhuang2019}, $L_{\text{[OII]}}=0.109^{+0.016}_{-0.006}\times L_{\text{[OIII]}}$ (their Equation 6). Using this relation, we get a maximum L(\oiii)$=6.1\times10^{43} \textrm{ erg s}^{-1}$. We can then employ the bolometric correction derived by \citet{lamastra2009} (454 for luminosities of $10^{42}-10^{44}\textrm{ erg s}^{-1}$) for \oiii to determine that the maximum bolometric luminosity of a quasar with EW(\oii)$\leq3$ will be $2.7\times10^{46} \textrm{ erg s}^{-1}$. We note that the real upper limit should be much lower, as only $6-12$\% of the \oii emission in quasars is expected to come from the AGN \citep{zhuang2019,wu2026}; this implies a limit closer to $\sim10^{45} \textrm{ erg s}^{-1}$, consistent with the maximum recovered luminosity of post-starburst quasars in our sample.

This result implies that post-starburst galaxy surveys are missing a significant population of post-starburst galaxies hosting luminous AGN and quasars, even when employing spectroscopic selection methods that allow for a significant quasar contribution. Based on both our H$\delta_{\text{A}}$ and \oii cuts, we can say that we definitely exclude quasars with bolometric luminosities in excess of $10^{46} \textrm{erg s}^{-1}$. Based on the allowed bolometric luminosity range from our H$\delta_{\text{A}}$ and EW(\oii) limits as well as the luminosity range of our sample ($44.0 \lesssim \textrm{Log(L}_{\text{Bol}}/\textrm{erg s}^{-1}) \lesssim 45.3$; see Table \ref{tab:psb_qsos}), we can expect that quasars with bolometric luminosities $\sim10^{45} \textrm{ erg s}^{-1}$ are excluded from our post-starburst selection. This luminosity range corresponds to a maximum recoverable black hole accretion rates of a few tenths of a solar mass per year and is at maximum a few percent of the Eddington luminosity of black holes in our mass range. 85\% of the SDSS DR16 quasar catalog has bolometric luminosities in excess of $\sim10^{45} \textrm{ erg s}^{-1}$ \citep{wu2022}, meaning we are missing a majority of the quasar population in post-starbursts. Importantly, as these objects are accreting at relatively low Eddington rates, we are missing the population of quasars that can drive the outflows that clear the environment around the quasar and potentially contribute to AGN feedback \citep{kingandpounds2015}. Post-starbursts with more luminous quasars can most effectively be identified via spectral \citep[e.g.]{krishna2025} or image \citep[e.g.][]{zhuang2024,sun2026} decomposition. New post-starburst identification techniques that allow for unobscured AGN will therefore be necessary to understand the full population of rapidly-quenching galaxies and test AGN feedback.

\subsection{Are post-starburst quasars fading?}
\label{discussion:fading}

\begin{figure}
    \centering
    \includegraphics[width=\linewidth]{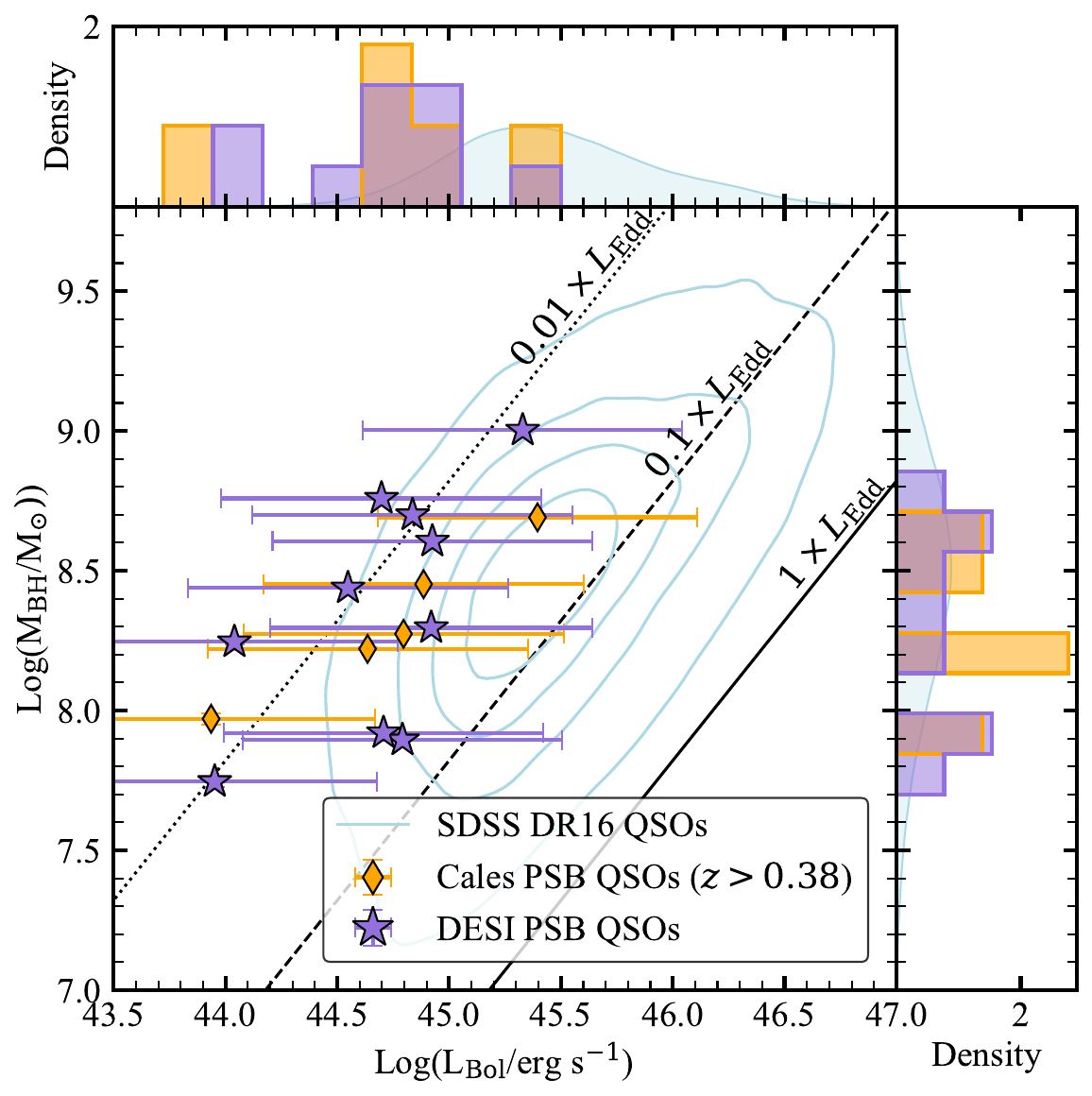}
    \caption{Black hole masses and bolometric luminosities for the DESI post-starburst quasars (purple stars) compared to the \citet[]{cales2013} post-starburst quasars at $z>0.38$ (yellow diamonds) and the SDSS quasar catalog at $0.49\leq z \leq1.39$ \citep[blue contours;][]{wu2022}. Lines indicating $1\%$, $10\%$, and $100\%$ of the Eddington luminosity at different black hole masses are indicated in dotted, dashed, and solid lines, respectively. DESI post-starburst quasars have similar black hole masses and AGN luminosities to their low-redshift counterparts but are lower in luminosity than the SDSS quasar catalog. Post-starburst quasars accrete at $1\% \lesssim \text{L}_{\text{Edd}} \lesssim 10\%$, likely due to the selection effects discussed in Section \ref{discussion:selectionmethods}.}
    \label{fig:mbh_ledd}
\end{figure}

\begin{figure} 
    \centering
    \includegraphics[width=\linewidth]{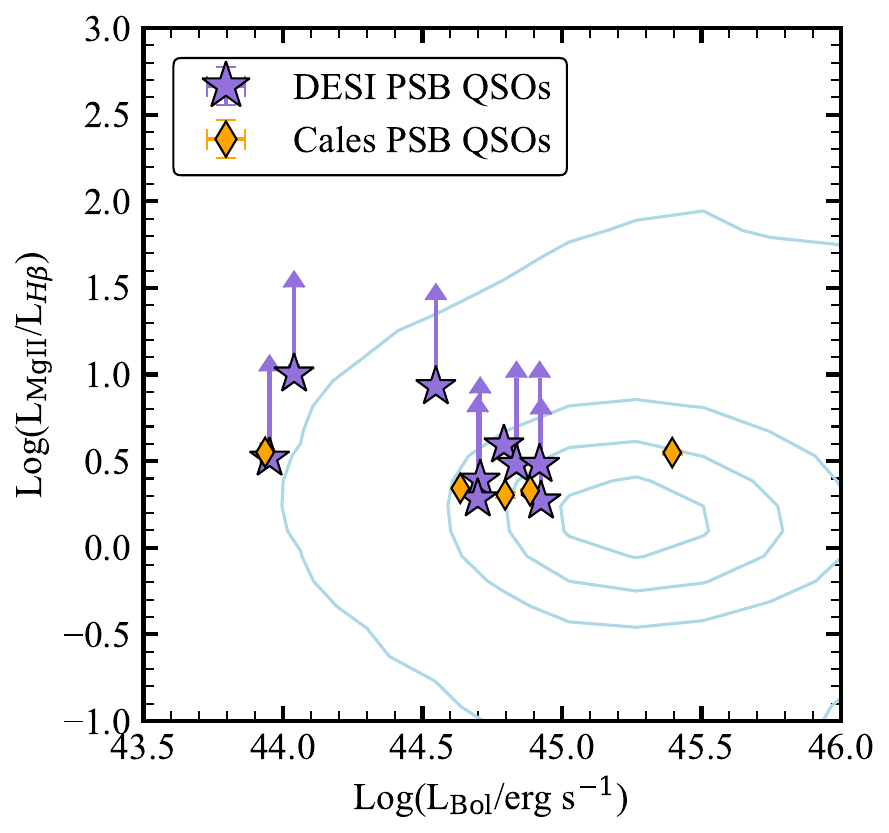}
    \caption{Bolometric luminosity versus the ratio of the luminosity of MgII to \hbeta  for the SDSS quasar catalog properties from \citet[]{wu2022} (light blue), the \citet[]{cales2013} post-starburst quasar sample at $0.38<z<0.42$ (orange), and the nine DESI post-starburst quasars from this work with spectral coverage of the \hbeta  rest-frame wavelength (purple). All but one of the post-starburst quasars have no detected broad \hbeta, and even with the lower limit on MgII/\hbeta, three of the post-starburst quasars fall outside $1\sigma$ of the SDSS MgII/\hbeta  distribution. These objects may be so-called ``MgII emitters," which have been theorized to be the remains of recently-faded quasars \citep[][]{guo2020}.}
    \label{fig:hbeta_mgii}
\end{figure}

As a result of the selection effects discussed above, the post-starburst quasars in our sample are systematically lower in bolometric luminosity than the SDSS quasars at the same redshift \citep{wu2022} and have similar luminosities and black hole masses to the low-redshift \citet{cales2011} sample at $z\geq0.38$. Post-starburst quasars at this redshift range are consistent with having Eddington ratios $0.01\lesssim\lambda \lesssim 0.1$, on the lower end of the SDSS quasar distribution (Figure \ref{fig:mbh_ledd}); this is likely due at least in part to the selection methods described in Section \ref{discussion:selectionmethods} and is again similar to the properties of the \citet{cales2011} sample. However, unlike post-starburst quasars at low redshift, we find that half of the post-starburst quasars in the \citet[]{soto2025} catalog with coverage of the \hbeta  emission line have no detected broad \hbeta  emission. 

A small fraction of AGN have been observed to have strong MgII emission relative to the AGN continuum and Balmer lines \citep[``MgII emitters,"][]{roig2014}. These objects were originally selected from the Baryon Oscillation Spectroscopic Survey \citep[BOSS;][]{boss} of the SDSS III \citep[][]{sdss3} at $0.35<z<1.1$. They have narrow line ratios indicative of AGN and tend to follow the \halpha/\hbeta  ratio distribution of other BOSS galaxies; only the ratio of MgII and NUV continuum to the Balmer lines is abnormal. Because they were initially selected from BOSS, the MgII emitters are also primarily identified in luminous red---and therefore, quiescent---galaxies. Several possible explanations for this phenomenon have been proposed, including that this is a more extreme example of the Baldwin effect, or the tendency in underluminous quasars for lower continuum luminosity to be associated with higher equivalent widths in emission lines, and vice versa \citep[][]{baldwin1977}. More recently, it has been proposed that these objects may be the fading remnants of more luminous quasars \citep[][]{guo2020}. This theory was motivated by the observation of several ``changing-look" AGN \citep[CL-AGN; see review by][]{ricci2023}, or AGN that gain or lose broad emission lines in the spectrum over the course of months or years, that retained their broad MgII emission after the broad Balmer lines had faded \citep[e.g.][]{macleod2016,yang2018,macleod2019,zeltyn2024}. Based on the discovery of a CL-AGN with fading MgII in \citet{guo2019}, \citet[]{guo2020} suggested that MgII should remain broad even after the continuum and broad \hbeta  have faded because MgII is emitted most efficiently further from the central engine and because it is intrinsically less responsive to changes in continuum than the Balmer lines. More recently, \citet{guo2025} used the DESI sample of CL-AGN to construct an evolutionary sequence in which broad \hbeta  fades first, followed by MgII and eventuall \halpha. This implies that MgII can remain in quasars for at least a few months \citep[the observed transition time for the object in][]{guo2019} even when the continuum luminosity has faded; based on observations of changing-look AGN with residual MgII, \citet{shen2021} estimated the timescale for MgII emitters to be $\Delta t_{\text{MgII}}\sim10$ years, much shorter than the timescales traced by EELRs \citep[$\sim10^{4-5}$ years; e.g.][]{lintott2009,keel2012,keel2017} and fossil outflows \citep[$\sim10^{6-7}$ years; e.g.][]{king2011,zubovas2018}.

In Figure \ref{fig:hbeta_mgii}, we compare MgII/broad \hbeta  ratio for our post-starburst quasar sample to the MgII/broad \hbeta  ratio for the \citet{cales2011} sample and the SDSS quasar catalog at $0.49\leq z\leq1.39$ \citep[][]{wu2022}. We restrict our analysis to the nine post-starburst quasars at $z<1.02$, as these objects have coverage of the \hbeta  line, and compare to the five objects in the \citet{cales2011} sample at $z>0.38$, for which the majority of the MgII line is included in the SDSS spectrum. Where no broad \hbeta  is detected, we report upper limits as described in Section \ref{sec:data}.

Only one of the nine DESI post-starburst quasars with \hbeta  coverage has a detected broad \hbeta  component. Of the five \citet{cales2011} objects with coverage of MgII, all have a detected broad \hbeta  component. MgII and \hbeta  luminosities and 2$\sigma$ upper limits are listed in Table \ref{tab:psb_qsos}, and the ratio of broad MgII to \hbeta  for both post-starburst quasar samples and for the SDSS DR16 quasars is shown in Figure \ref{fig:hbeta_mgii}. Three DESI post-starburst quasars fall outside the $1\sigma$ MgII/broad \hbeta  value for SDSS quasars at this redshift range, roughly a 2x overrepresentation. To test whether the post-starburst quasars have unusual MgII/\hbeta   ratios, we employ the log-rank survival test. We find that the DESI post-starburst quasars have a significantly different distribution of MgII/\hbeta   ratios than the SDSS DR16 quasars at similar redshift (test statistic 11.49, $p<0.005$).

Are our post-starburst quasars fading, or are they simply on the lower end of the quasar luminosity function? As discussed in Section \ref{discussion:selectionmethods}, our methods select against post-starbursts with AGN fractions $\gtrsim50\%$, creating a de facto upper limit on our recoverable AGN luminosities. To test whether this limit impacts our ability to identify post-starbursts with significant broad \hbeta, we test whether the ratio of MgII/\hbeta   correlates strongly with bolometric luminosity in the \citet[]{wu2022} SDSS quasar catalog. Under a Pearson R test, we detect a slight but significant correlation between bolometric luminosity and MgII/\hbeta  with $r=-0.051$ and $p<1\times10^{-5}$. It is therefore possible that the unusual MgII/\hbeta   ratios of our post-starburst quasar sample are due to the lower luminosity of this sample, causing the broad component of \hbeta   to be too faint to be detected in a majority of objects.   

The relatively low rate of MgII emitters makes it unlikely that our post-starburst quasars are fading CL-AGN. \citet{roig2014} finds that $\sim0.1\%$ of luminous BOSS galaxies at $0.35 \leq z \leq 1.1$ are MgII emitters, and \citet{guo2019} finds that $0.02\%$ of SDSS galaxies and quasars at $0.4\leq z \leq 0.8$ are MgII emitters. There are 8,786 DESI Early Data Release galaxies at $0.49 \leq z \leq 1.39$ that pass our SNR cut, meaning there should be $\sim2-10$ total MgII emitters in DESI at this redshift range. $\sim4\%$ of DESI Early Data Release galaxies in this redshift range are post-starburst, meaning at maximum we should have $\sim0.4$ post-starburst MgII emitters in our sample. This assumes that the rate of MgII emitters is constant in all galaxy types, which does not in principle need to be true; however, if the nine objects that completely lack \hbeta  emission are MgII emitters, this would imply an overrepresentation of $\sim20$x during the post-starburst phase. There is no evidence that such a dramatic overrepresentation exists for CL-AGN: while some studies have found that CL-AGN are more common in galaxies shutting down star formation \citep[e.g.][]{dodd2021,liu2021} and/or post-merger galaxies \citep[e.g.][]{charlton2019,tian2026}, others find no difference between the host galaxies of CL-AGN and other AGN \citep[e.g.][]{charlton2019,yu2020,verrico2025,agrawal2025,tian2026,zeltyn2026}, and no study has found a link between CL-AGN and specifically post-starburst galaxies. On the other hand, CL-AGN transitions occur primarily in objects with $0.01 \lesssim \lambda \lesssim 0.1$ \citep[e.g.][]{jana2024,zeltyn2024,guo2024b} which is consistent with the properties of our sample; it is possible that the same selection bias that removes luminous quasars from post-starburst samples is selecting for a higher fraction of CL-AGN and MgII emitters than among the broader galaxy population.

To confirm or rule out that these objects are MgII emitters and/or faded CL-AGN, we would need to include an analysis of the flux in \halpha, as it is much more luminous than \hbeta. None of the DESI spectra for our post-starburst quasar sample cover \halpha. For the lowest-redshift object in our sample (J1107+5244, $z=0.53$), there is an existing SDSS spectrum with \halpha coverage; however, as the maximum rest-frame wavelength covered by the SDSS spectrum is $6674\textrm{\AA}$, \halpha is in an extremely noisy part of the spectrum, and we are therefore unable to detect a broad \halpha component. Continued monitoring and more sensitive coverage of the region surrounding \halpha is necessary to determine whether post-starburst quasars in our sample are fading or whether they are simply lower in luminosity than the SDSS quasar sample.

\section{Conclusions}

In this work, we present ten post-starburst quasars at $0.49\leq z \leq1.39$ selected from the \citet[]{soto2025} sample of post-starburst galaxies in the DESI Early Data Release \citep[][]{desi_edr}. This is the first sample of post-starburst quasars to be selected from a post-starburst galaxy parent sample rather than a quasar catalog. These post-starburst quasars cover the crucial redshift range between low-redshift post-starburst quasars \citep[][]{brotherton1999,cales2011,melnick2015,wei2018} and the small samples of post-starburst quasars at high redshift \citep[e.g.][]{onoue2025,valentino2026}, tracing the decline of the cosmic star formation rate when most of the massive galaxies in the local universe quenched.

We find:

\begin{enumerate}
    \item 10/222 (\qsorate) post-starburst galaxies in the \citetalias{soto2025} catalog host unobscured AGN using a post-starburst quasar selection of $L_{\text{MgII}} > 3 \times 10^{41}$ erg s$^{-1}$. These objects have bolometric luminosities $44.0 \leq \textrm{Log(L}_{\text{Bol}}/\textrm{erg s}^{-1}) \leq 45.3$. This is a systematically lower luminosity range than that of SDSS quasars at matched redshift.
    \item Using a range of multiwavelength AGN indicators, we find that 47/222 ($26^{+3.0}_{-2.8}\%$) of \citetalias{soto2025} post-starbursts host AGN. This includes a partially overlapping set of 10 post-starburst quasars, $18-27$ narrow-line AGN, $1-3$ radio AGN, $6-8$ MIR AGN, and one X-ray AGN. 
    \item The presence of AGN contamination in a post-starburst spectrum removes the object from typical photometric, spectroscopic, and star-formation-based post-starburst selections once the AGN reaches a luminosity of $\textrm{L}_{\text{Bol}}\sim10^{45} \textrm{ erg s}^{-1}$, consistent with the maximum luminosity of unobscured AGN in our sample. This excludes a significant population of AGN at high accretion rates in post-starbursts, which may be the population of objects responsible for gas removal early in quenching.
    \item 9/10 post-starburst quasars in our sample have no detected broad \hbeta  component. These objects are similar to the ``MgII emitters" identified by \cite{roig2014} and may represent a population of fading changing-look AGN. The relative rarity of MgII emitters makes it more likely that our post-starburst quasars are simply lower in luminosity than quasars at the same redshift range. If they are fading quasars, this would represent a $\sim20$x overrepresentation of changing-look AGN during the post-starburst phase.
\end{enumerate}

These results highlight the importance of allowing for significant AGN contribution in the spectra of post-starburst galaxies when developing post-starburst selection techniques. Quasar episodes during the post-starburst phase may be crucial to drive outflows that remove molecular gas from galaxies; on the other hand, these AGN may simply be ``along for the ride," fueled by residual gas as star formation fades. Measuring the rate of luminous AGN throughout the post-starburst phase will require the development of spectral decomposition techniques that can be used to infer galaxy properties in the presence of significant AGN contamination; in an upcoming work, we present SED fits to our post-starburst quasar sample that account for AGN contamination in the UV/optical to determine whether the quasar rate changes after the starburst (Verrico et al, in preparation). At the same time, these techniques will need to be applied to large surveys to identify a statistical sample of AGN at all wavelengths during the brief post-starburst phase. Next-generation spectroscopic surveys like the Roman High Latitude Wide Area Survey and Subaru's Prime Focus Spectrograph survey will be critical to understand whether AGN feedback plays an important role in rapid galaxy quenching.

\begin{acknowledgments}
M.E.V. and K.D.F would like to acknowledge funding from NSF Grant AST-2307375. 

M.E.V. would like to thank Yue Shen, Benny Trakhtenbrot, Qiaoya Wu, Grisha Zeltyn, and Ming-Yang Zhuang for conversations that improved this work. M.E.V. would also like to thank Nicholas Earl for code used in the preparation of this work.

This research used data obtained with the Dark Energy Spectroscopic Instrument (DESI). DESI construction and operations is managed by the Lawrence Berkeley National Laboratory. This material is based upon work supported by the U.S. Department of Energy, Office of Science, Office of High-Energy Physics, under Contract No. DE–AC02–05CH11231, and by the National Energy Research Scientific Computing Center, a DOE Office of Science User Facility under the same contract. Additional support for DESI was provided by the U.S. National Science Foundation (NSF), Division of Astronomical Sciences under Contract No. AST-0950945 to the NSF’s National Optical-Infrared Astronomy Research Laboratory; the Science and Technology Facilities Council of the United Kingdom; the Gordon and Betty Moore Foundation; the Heising-Simons Foundation; the French Alternative Energies and Atomic Energy Commission (CEA); the National Council of Humanities, Science and Technology of Mexico (CONAHCYT); the Ministry of Science and Innovation of Spain (MICINN), and by the DESI Member Institutions: www.desi.lbl.gov/collaborating-institutions. The DESI collaboration is honored to be permitted to conduct scientific research on I’oligam Du’ag (Kitt Peak), a mountain with particular significance to the Tohono O’odham Nation. Any opinions, findings, and conclusions or recommendations expressed in this material are those of the author(s) and do not necessarily reflect the views of the U.S. National Science Foundation, the U.S. Department of Energy, or any of the listed funding agencies.

This work is based on data from eROSITA, the soft X-ray instrument aboard SRG, a joint Russian-German science mission supported by the Russian Space Agency (Roskosmos), in the interests of the Russian Academy of Sciences represented by its Space Research Institute (IKI), and the Deutsches Zentrum für Luft- und Raumfahrt (DLR). The SRG spacecraft was built by Lavochkin Association (NPOL) and its subcontractors, and is operated by NPOL with support from the Max Planck Institute for Extraterrestrial Physics (MPE). The development and construction of the eROSITA X-ray instrument was led by MPE, with contributions from the Dr. Karl Remeis Observatory Bamberg \& ECAP (FAU Erlangen-Nuernberg), the University of Hamburg Observatory, the Leibniz Institute for Astrophysics Potsdam (AIP), and the Institute for Astronomy and Astrophysics of the University of Tübingen, with the support of DLR and the Max Planck Society. The Argelander Institute for Astronomy of the University of Bonn and the Ludwig Maximilians Universität Munich also participated in the science preparation for eROSITA. 

The National Radio Astronomy Observatory is a facility of the National Science Foundation operated under cooperative agreement by Associated Universities, Inc. CIRADA is funded by a grant from the Canada Foundation for Innovation 2017 Innovation Fund (Project 35999), as well as by the Provinces of Ontario, British Columbia, Alberta, Manitoba and Quebec. 

Funding for the Sloan Digital Sky Survey V has been provided by the Alfred P. Sloan Foundation, the Heising-Simons Foundation, the National Science Foundation, and the Participating Institutions. SDSS acknowledges support and resources from the Center for High-Performance Computing at the University of Utah. SDSS telescopes are located at Apache Point Observatory, funded by the Astrophysical Research Consortium and operated by New Mexico State University, and at Las Campanas Observatory, operated by the Carnegie Institution for Science. The SDSS web site is \url{www.sdss.org}.

SDSS is managed by the Astrophysical Research Consortium for the Participating Institutions of the SDSS Collaboration, including the Carnegie Institution for Science, Chilean National Time Allocation Committee (CNTAC) ratified researchers, Caltech, the Gotham Participation Group, Harvard University, Heidelberg University, The Flatiron Institute, The Johns Hopkins University, L'Ecole polytechnique f\'{e}d\'{e}rale de Lausanne (EPFL), Leibniz-Institut f\"{u}r Astrophysik Potsdam (AIP), Max-Planck-Institut f\"{u}r Astronomie (MPIA Heidelberg), Max-Planck-Institut f\"{u}r Extraterrestrische Physik (MPE), Nanjing University, National Astronomical Observatories of China (NAOC), New Mexico State University, The Ohio State University, Pennsylvania State University, Smithsonian Astrophysical Observatory, Space Telescope Science Institute (STScI), the Stellar Astrophysics Participation Group, Universidad Nacional Aut\'{o}noma de M\'{e}xico, University of Arizona, University of Colorado Boulder, University of Illinois at Urbana-Champaign, University of Toronto, University of Utah, University of Virginia, Yale University, and Yunnan University.  

This publication makes use of data products from the Wide-field Infrared Survey Explorer, which is a joint project of the University of California, Los Angeles, and the Jet Propulsion Laboratory/California Institute of Technology, funded by the National Aeronautics and Space Administration.
\end{acknowledgments}

\software{Astropy \citep{astropy:2013,astropy:2018,astropy:2022}, Prospector \citep{johnson17,leja17,johnson20}, PyQSOFit \citep{guo2018, shen2019}, SEDpy \citep{sedpy}.
          }

\bibliography{sample701}{}
\bibliographystyle{aasjournalv7}

\newpage

\appendix 
\vspace{-10pt}
\section{\texttt{Prospector} Stellar Population Modeling }\label{appendix:prospector_model}

We use \texttt{Prospector} stellar population fitting software \citep[][]{johnson17,leja17,johnson2021} to compute the redshifts and stellar mass estimates used throughout this work. \texttt{Prospector} is a stellar population synthesis software based on FSPS \citep[][]{conroy2009, conroy2010} accessed through py-FSPS \citep[][]{pyfsps} that forward models galaxy spectra and photometry to determine galaxy properties. We use the MILES spectral library \citep{sanchez-blazquez2006} and the MIST isochrones \citep{dotter2016, choi2016, paxton2011, paxton2013, paxton2015} in our fitting. We use the \texttt{continuity\_psb\_sfh} star formation template developed by \citet[]{suess2022b} which is optimized for post-starburst galaxies but which has also been tested on star-forming and quiescent galaxies. This template was used to fit the \squiggle post-starburst galaxies at similar redshift to our sample \citep[][]{suess2022a}.  

Stellar population fitting in AGN hosts is difficult, as the blue continuum and emission lines from AGN can easily be confused with star formation and vice versa. We address this challenge and present the full results of our stellar population fitting in an upcoming work (Verrico et al. in preparation). For the purposes of this work, we use the \texttt{agn\_bbb} model from \citet[]{bingjiewang2025} to account for AGN contamination. This model contains a big blue bump model with a free parameter $f_{\text{AGN}} \equiv \frac{F_{\nu, \text{AGN, 5500 \AA}}}{F_{\nu, \text{Host, 5500 \AA}}}$. We integrate a prior informed by our results from PyQSOFit (Normal($f_{\text{AGN, PyQSOFit}}$, $\sigma=0.1$). We compute $f_{\text{AGN, PyQSOFit}}$ using the recovered host fraction from PyQSOFit at 4200 \AA due to the lack of coverage at 5100 \AA. For the one object for which QSO-host decomposition fails, we center the prior at $f_{\text{AGN}} = 0$. We present these parameters and the others used for our stellar population fitting in Table \ref{tab:prospector} below.

\begin{table}[h]
    \centering
    \begin{tabular}{c|c|c|c}
         Name & Description & Prior & (Initial/Central) Value \\
         \hline
         zred & Redshift & Normal, $\sigma=5\times10^{-3}$ & DESI Value \\
         dust\_type & Dust attenuation model & Fixed & \citet[]{kriek2013} \\
         dust2 & Dust attenuation & Uniform[0.0, (Log($e$))$^{-1}$] & (5$\times$Log($e$))$^{-1}$ \\
        dust1 & Dust attenuation around stellar birth clouds & Fixed & $2\times$dust2 \\
        duste\_umin & Minimum radiation field strength & Uniform[0.1, 25] & $1.0$ \\
        duste\_gamma & Relative dust heating at $U_{\text{min}}$ & Uniform[$1\times10^{-4}$,1] & $1\times10^{-2}$ \\
        duste\_qpah & PAH grain size distribution & Uniform[$1\times10^{-2}$, 10] & $2.0$ \\
         imftype & Initial mass function & Fixed & \citet[]{chabrier2003} \\
         logmass & Mass formed & Uniform[7,12] & \citet[]{bell2003} Table 7 \\
         logzsol & Metallicity & Normal & \citet[][]{gallazzi2005} Table 2 \\
         sfrratio\_old & SFR in old bins to the first flex bin & Student T & [0.0, 0.0, 0.0] \\
        sfrratio\_young & SFR in the youngest bin to the last flex bin & Student T & 0.0 \\
        sfrratio & SFR ratios in flex bins & Student T & [0.0, 0.0, 0.0, 0.0] \\
        tflex & The length of time containing all flex bins & Fixed & 2 Gyr \\
        tlast & The length of the last time bin & Uniform [0.01, 1.99] & 1 Gyr \\
        f\_outlier\_spec & Fraction of outlier pixels & Uniform [0.0, 0.1] & 0.001 \\
        spec\_jitter & Spectroscopic white noise term & Uniform [0.5, 15.0] & 1.0 \\
        sigma\_smooth & Spectral smoothing  & Normal & Fuji VDISP \\
        fagn\_bbb$^*$ & F$_{\nu, \text{AGN}}$/F$_{\nu, *}$ at rest 5500\AA & Normal, $\sigma=0.1$ & $1-$PyQSOFit host fraction \\
        fagn$^*$ & L$_{\text{Bol, AGN}}$/L$_{\text{Bol,*}}$ & Fixed & $1\times10^{-4}$ \\
        agn\_tau$^*$ & Optical depth of torus clouds & Fixed & 5.0 \\
    \end{tabular}
    \caption{Parameters used in our \texttt{Prospector} fits. Only the resulting redshift (zred) and stellar mass (logmass, corrected for the fraction of stars surviving at observation) are used in this work. *The fagn\_bbb parameter controls the normalization of the AGN big blue bump model introduced by \citet[]{bingjiewang2025}, while the fagn and agn\_tau parameters control the MIR AGN model described in \citet[]{leja2018}. These two fractions are not tied to each other in any way. We include AGN dust for the \citet[]{leja2018} MIR AGN model, but we do not incorporate the dust4 parameter described in \citet[]{bingjiewang2025}.}
    \label{tab:prospector}
\end{table}

\section{Multiwavelength AGN properties of the S25 catalog }\label{appendix:agn}

Here, we include a table of the multiwavelength AGN properties of individual objects in the \citetalias{soto2025} sample. A portion of the table is reproduced here; the full table is available in the journal version of this paper. Note that for FIRST and VLASS, we indicate whether the object is detected rather than selected as an AGN. We demonstrate in Section \ref{sec:radio} that J1058+3314 is likely an AGN, while J1409-0139 and J1611+4415 likely have at least some contribution from star formation.

\begin{sidewaystable}[h]
    \centering
        \begin{tabular}{c|c|c|c|c|c|c|c|c|c|c|c|c|c}
        Name & DESI ID & RA & Dec & Z & MgII & MEx & MEx $>3$ & OHNO & \nev & VLASS & FIRST & WISE & X-Ray \\
         & &   &  &  & Selected? & Selected? & Selected? & Selected? & Detected? & Detected? & Detected? & Selected? & Selected? \\
        \hline
         J0549-2424 & 39627211644342342 & 87.36231 & -24.41017 & 0.49 & 0.0 & 0.0 & 0.0 & 1.0 & 0.0 & 0.0 & 3.0 & 1.0 & 0.0 \\
        J0651+3839 & 39633052476903349 & 102.89483 & 38.66234 & 0.49 & 0.0 & 0.0 & 0.0 & 0.0 & 0.0 & 0.0 & 3.0 & 0.0 &  $-$ \\
        J1809+6219 & 39633414176903213 & 272.32551 & 62.31797 & 0.49 & 0.0 & 0.0 & 0.0 & 1.0 & 0.0 & 0.0 & 3.0 & 0.0 &  $-$ \\
        J1243+6058 & 39633399110962714 & 190.75493 & 60.97054 & 0.49 & 0.0 & 0.0 & 0.0 & 0.0 & 0.0 & 0.0 & 0.0 & 0.0 &  $-$ \\
        J1609+5506 & 39633322476833246 & 242.28618 & 55.11198 & 0.49 & 0.0 & 0.0 & 0.0 & 0.0 & 0.0 & 0.0 & 0.0 & 0.0 &  $-$ \\
        J0845+2325 & 39628339576574809 & 131.48869 & 23.42167 & 0.5 & 0.0 & 0.0 & 0.0 & 0.0 & 0.0 & 0.0 & 0.0 & 0.0 & 0.0 \\
        J0955+3721 & 39633024433783417 & 148.75242 & 37.36328 & 0.5 & 0.0 & 0.0 & 0.0 & 0.0 & 0.0 & 0.0 & 0.0 & 0.0 & 0.0 \\
        J1059+3208 & 39628527137458358 & 164.79714 & 32.14354 & 0.5 & 0.0 & 0.0 & 0.0 & 0.0 & 0.0 & 0.0 & 0.0 & 0.0 & 0.0 \\
        J0838-0544 & 39627648137170017 & 129.67876 & -5.74817 & 0.5 & 0.0 & 0.0 & 0.0 & 0.0 & 0.0 & 0.0 & 3.0 & 0.0 & 0.0 \\
        J0932+8151 & 39633559501146261 & 143.11837 & 81.86353 & 0.5 & 0.0 & 0.0 & 0.0 & 0.0 & 0.0 & 0.0 & 3.0 & 0.0 &  $-$ \\
        J1303+2654 & 39628417167001682 & 195.79457 & 26.90247 & 0.5 & 0.0 & 0.0 & 0.0 & 0.0 & 0.0 & 0.0 & 0.0 & 0.0 &  $-$ \\
        J0222-0412 & 39627682689847304 & 35.64701 & -4.21074 & 0.5 & 0.0 & 0.0 & 0.0 & 1.0 & 0.0 & 0.0 & 0.0 & 0.0 &  $-$ \\
        J0234+2638 & 39628409411732464 & 38.62314 & 26.63407 & 0.5 & 0.0 & 0.0 & 0.0 & 0.0 & 0.0 & 0.0 & 3.0 & 0.0 &  $-$ \\
        J0914+8420 & 39633567050891506 & 138.72949 & 84.33907 & 0.5 & 0.0 & 0.0 & 0.0 & 0.0 & 0.0 & 0.0 & 3.0 & 0.0 &  $-$ \\
        J1613+5346 & 39633304843978463 & 243.27613 & 53.76966 & 0.51 & 0.0 & 0.0 & 0.0 & 0.0 & 0.0 & 0.0 & 0.0 & 0.0 &  $-$ \\
        J0709+5658 & 39633348347301955 & 107.31317 & 56.96914 & 0.51 & 0.0 & 0.0 & 0.0 & 0.0 & 0.0 & 0.0 & 0.0 & 0.0 &  $-$ \\
        J0628+4407 & 39633147909902078 & 97.13668 & 44.12176 & 0.51 & 0.0 & 0.0 & 0.0 & 0.0 & 0.0 & 0.0 & 3.0 & 0.0 &  $-$ \\
        J1108+3252 & 39632940514151664 & 167.14067 & 32.87836 & 0.51 & 0.0 & 0.0 & 0.0 & 1.0 & 0.0 & 0.0 & 0.0 & 0.0 & 0.0 \\
        J1013+3210 & 39628526978072716 & 153.44919 & 32.17858 & 0.51 & 0.0 & 0.0 & 0.0 & 0.0 & 0.0 & 0.0 & 0.0 & 0.0 & 0.0 \\
        J1750+6154 & 39633411307997723 & 267.66595 & 61.91157 & 0.51 & 0.0 & 0.0 & 0.0 & 0.0 & 0.0 & 0.0 & 3.0 & 0.0 &  $-$ \\
        J1204-0022 & 39627775685953869 & 181.11408 & -0.37939 & 0.51 & 0.0 & 0.0 & 0.0 & 0.0 & 0.0 & 0.0 & 0.0 & 0.0 & 0.0 \\
        J0940+6416 & 39633435077116405 & 145.22325 & 64.27135 & 0.52 & 0.0 & 0.0 & 0.0 & 0.0 & 0.0 & 0.0 & 0.0 &  $-$ &  $-$ \\
        J1615+5406 & 39633308425912965 & 243.9981 & 54.10177 & 0.52 & 0.0 & 0.0 & 0.0 & 0.0 & 0.0 & 0.0 & 0.0 & 0.0 &  $-$ \\
        J1551+4247 & 39633127609467377 & 237.77454 & 42.78548 & 0.52 & 0.0 & 0.0 & 0.0 & 0.0 & 0.0 & 0.0 & 0.0 & 0.0 &  $-$ \\
        J1245+6207 & 39633410704016351 & 191.26527 & 62.11904 & 0.52 & 0.0 & 0.0 & 0.0 & 0.0 & 0.0 & 0.0 & 0.0 &  $-$ &  $-$ \\
        J1800+6108 & 39633402688701224 & 270.09475 & 61.14793 & 0.52 & 0.0 & 0.0 & 0.0 & 0.0 & 0.0 & 0.0 & 3.0 & 0.0 &  $-$ \\
        J0701+5622 & 39633338301940288 & 105.27418 & 56.36884 & 0.52 & 0.0 & 0.0 & 0.0 & 0.0 & 0.0 & 0.0 & 0.0 & 0.0 &  $-$ \\
        J1751+6130 & 39633405570188405 & 267.78277 & 61.51628 & 0.52 & 0.0 & 0.0 & 0.0 & 0.0 & 0.0 & 0.0 & 3.0 & 0.0 &  $-$ \\
        J0423-0437 & 39627677174334909 & 65.99778 & -4.62326 & 0.52 & 0.0 & 0.0 & 0.0 & 0.0 & 0.0 & 0.0 & 3.0 &  $-$ & 0.0 \\
        J1219-0040 & 39627769709071008 & 184.88272 & -0.67087 & 0.53 & 0.0 & 0.0 & 0.0 & 0.0 & 0.0 & 0.0 & 0.0 & 0.0 & 0.0 \\
        \end{tabular}
        \caption{Multiwavelength AGN properties for the \citetalias{soto2025} post-starburst sample. AGN selection criteria are described in Section \ref{sec:results}. Columns with missing data reflect objects with insufficient wavelength coverage to perform the relevant AGN selection. A portion of the table is shown here for form and content; the full machine-readable table is linked in the online version.}
\end{sidewaystable}

\end{document}